\documentclass[paper=A4,11pt,DIV=12,headings=big]{scrartcl}
\pdfoutput=1

\usepackage[a4paper, top=1.2in, bottom=1.2in, left=1.1in, right=1.1in]{geometry}
\usepackage{amsfonts}
\usepackage[normalem]{ulem}
\usepackage{amsmath,amssymb,slashed}
\usepackage[normal]{caption}

\usepackage{cancel}
\usepackage{xcolor}
\usepackage{cite,bbm}
\usepackage{enumerate}
\usepackage{graphicx}
\graphicspath{{./Figs/}}
\usepackage{wrapfig}

\renewenvironment{abstract}{%
\begin{minipage}{0.95\textwidth}
}
{\par\noindent\end{minipage}}

\usepackage[multiple]{footmisc}
\let\oldfootnote\footnote\renewcommand\footnote[1]{\oldfootnote{\hspace{2mm}#1}}

\definecolor{darkblue}{rgb}{0,0,0.9}
\usepackage{hyperref}
\hypersetup{
linktocpage,
colorlinks,
citecolor=darkblue,
filecolor=darkblue,
linkcolor=darkblue,
urlcolor=darkblue,
pdftitle={Lab-frame observables for probing the top-Higgs interaction},%
pdfauthor={}
}
\allowdisplaybreaks
\addtokomafont{disposition}{\rmfamily\boldmath}
\newcommand{\mailref}[1]{\href{mailto:#1}{#1}}

\def\sla#1{\setbox0=\hbox{$#1$}\dimen0=\wd0
      \setbox1=\hbox{/} \dimen1=\wd1 \ifdim\dimen0>\dimen1
      \rlap{\hbox to \dimen0{\hfil/\hfil}} #1                        \else
      \rlap{\hbox to \dimen1{\hfil$#1$\hfil}}
      /   \fi}

\newcommand{\be}{\begin{equation}}
\newcommand{\ee}{\end{equation}}
\newcommand{\bea}{\begin{eqnarray}}
\newcommand{\eea}{\end{eqnarray}}

\newcommand{\nn}{\nonumber}

\def \tth{t\bar{t}h}
\def \ttb{t\bar{t}}

\def \gamgam{\gamma \gamma}
\def \Rg[#1]{\frac{\Gamma(h\to #1)}{\Gamma(h \to #1)^{\rm SM} }}
\def \sigh[#1]{\frac{\sigma( #1)}{\sigma( #1)^{\rm SM} }}
\newcommand{\Lag}{\mathcal L}
\def\Lag{\mathcal{L}}
\newcommand{\at}{a_t}
\newcommand{\bt}{b_t}

\newcommand{\mc}{\mathcal}
\def\cw{\cos\theta_{w}}

\def \pt {p_{T}}
\newcommand{\sgn}{{\rm sgn}}

\newcommand{\dhww}{h\rightarrow W^{(*)}W^{(*)}}
\newcommand{\dhzz}{h\rightarrow Z^{(*)}Z^{(*)}}
\begin{document}

%%%% TITLE PAGE
\begin{flushright}
\small
LAPTH-004/15\\
MSUHEP-150113\\
%CERN-PH-TH/2013-190
\end{flushright}
\vskip0.5cm

%\begin{flushright}
%Last modified on \today
%\end{flushright}

\begin{center}
%%%% TITLE
{\sffamily \bfseries \LARGE \boldmath
Lab-frame observables for probing the top-Higgs interaction}\\[0.8 cm]
%%%% AUTHORS
{\normalsize \sffamily \bfseries Fawzi~Boudjema$^a$, Rohini~M.~Godbole$^b$, Diego~Guadagnoli$^a$, Kirtimaan~A.~Mohan$^c$} \
\\[0.5 cm]
\small
$^a$ { LAPTh, Universit\'e de Savoie et CNRS, BP110, F-74941 Annecy-le-Vieux Cedex, France}\\[0.1cm]
$^b${ CHEP, Indian Institute of Science, Bangalore 560012, India}\\[0.1cm]
$^c${ Department of Physics and Astronomy, Michigan State University, East Lansing, Michigan 48824, USA}\\[0.4cm]
{ E-mail:} \mailref{fawzi.boudjema@lapth.cnrs.fr}, \mailref{rohini@cts.iisc.ernet.in}, 
\mailref{diego.guadagnoli@lapth.cnrs.fr}, \mailref{kirtimaan@pa.msu.edu}%
\end{center}

\medskip

\begin{abstract}
\noindent
We investigate methods to explore the CP nature of the $\tth$ coupling at the LHC, focusing on 
associated production of the Higgs with a $\ttb$ pair.
We first discuss the constraints implied by low-energy observables and by the Higgs-rate information 
from available LHC data, emphasizing that they cannot provide conclusive evidence on the
nature of this coupling. We then investigate kinematic observables that could probe the $\tth$ 
coupling directly, in particular quantities that can be constructed out of just {\em lab-frame} 
kinematics. 
We define one such observable by exploiting the fact that $\ttb$ spin correlations do 
also carry information about the CP-nature of the $\tth$ coupling. 
Finally, we introduce a CP-odd quantity and a related asymmetry, able to probe CP violation in 
the $\tth$ coupling and likewise constructed out of lab-frame momenta only.
\end{abstract}

\vspace{1.0cm}

%%%% TOC
\setcounter{tocdepth}{2}
\noindent \rule{\textwidth}{0.3pt}\vspace{-0.4cm}\tableofcontents
\noindent \rule{\textwidth}{0.3pt}
%%%% END TOC

\renewcommand{\thefootnote}{\arabic{footnote}}
\setcounter{footnote}{0}
%%%% END TITLE PAGE

\section{Introduction}

The 7-8 TeV runs of the LHC have led to the discovery of a scalar particle with a mass $m_h \simeq $ 125 
GeV \cite{Aad:2012tfa,Chatrchyan:2012ufa,Chatrchyan:2013lba,Aad:2013xqa}. The properties measured so far 
show very good consistency with those expected for the Standard-Model (SM) Higgs boson.
Further, these runs have not revealed the existence of new particles. The fact remains that the SM cannot 
address a few pressing questions, such as the baryon asymmetry of the universe, the large mass hierarchy 
in the fermion sector as well as an explanation for the Dark Matter abundance in the Universe. These issues 
call for new physics (NP) beyond the SM. Furthermore, the observation of a 125 GeV {\it elementary} scalar, 
as well as the absence so far of NP at the TeV scale, leave unanswered the question of why its mass $m_h$ 
is so different than the gravitational scale.

In order to ease the explanation of the observed baryon asymmetry, new sources of CP violation are 
desirable. Such sources exist in many simple extensions of the SM. One notable example is an extended 
Higgs sector such as a two Higgs doublet model. 
Therein, CP violation is incorporated in the Higgs sector through mixing of CP-even and -odd states.
Within these models, the 125-GeV boson identified with the Higgs can have indefinite CP quantum numbers 
due to mixing of CP-even and -odd states. A determination of the CP nature of this particle and its 
interactions may thus hold clue of NP.

A program to probe the CP nature of the discovered Higgs scalar is already under way at the LHC 
experiments. The pure pseudo-scalar hypothesis has already been ruled out at greater than $95\%$ 
confidence level (CL) and consistency with the CP-even nature established by ATLAS 
\cite{ATLAS-CONF-2013-013} and CMS \cite{Chatrchyan:2013mxa,Chatrchyan:2012jja}. This has been achieved 
by an analysis of the $hZZ$ coupling using $(\dhzz)$ decay channel.%
\footnote{For discussions  of the $(\dhzz)$ decay mode as a probe of the Higgs CP properties see, for example, refs.~\cite{Choi:2002jk,Godbole:2007cn,Gao:2010qx,DeRujula:2010ys,Stolarski:2012ps,Bolognesi:2012mm,Chen:2014ona,Chatrchyan:2013iaa,Khachatryan:1969386,Dawson:2013bba}. It is also possible to probe the same in Vector Boson Fusion~\cite{Plehn:2001nj,Hankele:2006ma,Andersen:2010zx,Andersen:2008gc,Djouadi:2013yb,Englert:2012xt} production and associated (Vh) production~\cite{Han:2009ra,Christensen:2010pf,Desai:2011yj,Ellis:2012xd,Godbole:2014cfa,Godbole:2013saa,Delaunay:2013npa}.}
It should be noted however that tree-level coupling of the CP-odd component of the Higgs to gauge bosons 
is in fact not allowed and can only proceed through loops. Couplings between the Higgs CP-odd component 
and gauge bosons manifest themselves as operators of dimension six (or higher) in the language of 
effective Lagrangians. The effect of such operators is expected to be suppressed in comparison to 
tree-level interactions.

On the other hand, the CP-odd component of the Higgs couples to fermions at the tree level. As a result 
the Higgs-fermion couplings provide an unambiguous and more sensitive probe of a CP-mixed state compared 
to Higgs-gauge-boson couplings.%
\footnote{Note that the Higgs to di-photon decay proceeds through loop processes at LO (unlike decays to 
$W$ and $Z$ bosons), making it sensitive to the CP-odd component of the Higgs.}
It is possible to probe Higgs-fermion couplings by studying Higgs decays to fermions. Since these are 
two-body decays of a spin $0$ particle, the CP nature of the coupling is reflected in the spin correlation 
of the decay fermions. Luckily, the spin information of the $t, \tau$ is also reflected in the decay 
products of the same. This as well as their larger couplings, offers possibilities of probing the CP 
nature of the Higgs through an analysis of the $h \tau \bar \tau$ and $h t \bar t$ coupling. Analysis of 
this coupling using the $h \rightarrow \tau \bar \tau$ case has been shown to be quite  promising for this 
purpose~\cite{Berge:2014sra,Berge:2012wm,Berge:2011ij,Berge:2008wi}.
However, a measure of the strength of the coupling CP-odd component in the $h\tau\tau$ interaction does 
not automatically qualify as a measure of the same for other fermions, i.e. the CP-odd component may not 
couple to all fermions universally (as is the prediction in some NP models).
It therefore becomes important to be able to probe the CP nature of the Higgs in all its couplings. 
The largest of all such couplings, $ t \bar t h$, cannot be tested by direct decay, because 
$h \rightarrow t \bar t$ is not allowed. However, the large value of this coupling implies large 
production rates for associated production of the Higgs with a $t \bar t$ pair, and this mode therefore 
qualifies as the most direct probe of the Higgs-top coupling and of the CP nature of the Higgs.

Also from a more general perspective, it is well known that the coupling of the Higgs to the top quark, 
is of great relevance, theoretically and experimentally alike. On the theoretical side, the importance of 
the top Yukawa coupling follows from the fact that it is numerically very close to unity. Such a large 
value of the Yukawa coupling is suggestive of an active role of the top quark in the generation of the 
electroweak-symmetry breaking (EWSB) scale. As a matter of fact, the Higgs-top interaction has important 
consequences on spontaneous symmetry breaking within the SM -- notably, on vacuum stability arguments -- 
as well as beyond the SM -- where the top drives electroweak symmetry breaking in some scenarios. 
Most importantly, this coupling drives the main production channel at the LHC (gluon fusion), and also 
contributes  to the crucial decay of the Higgs into two photons.

The above considerations justify the importance of measuring the top-Higgs coupling with the highest 
accuracy achievable, and, in particular, of determining it by {\it direct measurement} via $\tth$ 
production. In this paper we focus on this possibility.

We parameterize the Higgs couplings to fermions through the effective Lagrangian 
\be
\Lag_{hf\bar{f}}= -\sum\limits_{f} \frac{m_f}{v} h \bar{f} (a_f+ i b_f\gamma_5)f, \label{eq:ffh}
\ee
where the sum is over all quarks and leptons.
In the SM, where the Higgs is a scalar, $a_f=1$ and $b_f=0$ for any fermion $f$. For a pure pseudo-scalar 
$a_f=0$ and $b_f \neq 0$. A Higgs with mixed CP properties is realized if both $a_f\neq0$ and $b_f\neq0$.
The exact values of these coefficients will depend on the specific model. Here we are interested in a 
model-independent approach to determine, from data, the nature of  the $t \bar t h$ interaction  which is 
potentially the largest coupling of all fermions.

The production and decay rates of the Higgs measured at the LHC 
\cite{Djouadi:2013qya,Cheung:2013kla,Cheung:2014oaa} do  provide important constraints on the strength of both 
$\at, \bt$. Indirect constraints will be discussed in more detail in Sec. \ref{sec:indirect_tth}, with the 
aim of spelling out the underlying assumptions that enter the derivations of these constraints. We will 
show that strong constraints on $\at$ and $\bt$ can be placed \textit{only} under these assumptions.

As argued, the most general and direct determination of the $\at,\bt$ couplings in eq. (\ref{eq:ffh}) is 
possible by measuring $\tth$ production.\footnote{%
An alternative approach, which we do not discuss in this work, to study the couplings in eq. 
(\ref{eq:ffh}), is to use single top production~\cite{Biswas:2012bd,Biswas:2013xva,Ellis:2013yxa,Yue:2014tya,Chang:2014rfa,Kobakhidze:2014gqa,Englert:2014pja,CMS-PAS-HIG-14-001,CMS-PAS-HIG-14-015,CMS-PAS-HIG-14-026}.} The 
$\tth$ production mode is notoriously hard to measure at the LHC, yet feasible. In fact, already with the 
limited data set of the 7 and 8 TeV runs of the LHC, the signal strengths in the $\tth$ production channel 
have been measured by both ATLAS~\cite{Aad:2014lma,ATLAS-CONF-2014-011} and CMS~\cite{Khachatryan:2014qaa}. 
Some preliminary studies suggest that a significant ($ > 5\sigma$) measurement of Higgs production in the 
$\tth$ channel is possible for upcoming runs of the LHC~\cite{ATL-PHYS-PUB-2014-012,Plehn:2009rk,Artoisenet:2013vfa,Buckley:2013auc,Maltoni:2002jr,Curtin:2013zua,Agrawal:2013owa}.

Needless to say, a measurement of the $\tth$ production cross-section alone is not sufficient to determine 
the vertex in eq.~(\ref{eq:ffh}) completely. To this end, it is necessary to consider in detail the $\tth$ 
production and the decay kinematics. In this paper we suggest and discuss useful discriminating observables 
to probe the vertex in eq. (\ref{eq:ffh}), with emphasis on those that can be defined directly in the lab 
frame. Note that, on the other hand, we refrain from entering the discussion about a precision 
determination of the vertex.

The rest of this paper is organized as follows. In sec. \ref{sec:indirect_tth} we describe and derive 
indirect constraints on the couplings $\at$ and $\bt$. In sec. \ref{sec:tth_obs} we then proceed to analyse 
$\tth$ production at the LHC and construct various observables, including a CP-violating one, that could be 
used to determine the nature of the $\tth$ interaction itself. Finally in sec. \ref{sec:conclusions} we 
summarize and conclude.

\section{Indirect probes of an anomalous $\tth$ coupling} \label{sec:indirect_tth}

Electric dipole moments (EDMs) can impose severe constraints on new CP-violating weak phases. 
A scalar with mixed parity that couples to both the electron and the top as described by 
eq.~(\ref{eq:ffh}) leads to CP violation through interference of the type $a_f b_{f^\prime}$. 
Indeed at 2-loop a Barr-Zee type diagram induces an EDM for the electron of the form 
$d_e \propto b_t a_e f_1(m_t^2 / m_h^2) + b_e a_t f_2(m_t^2 / m_h^2)$, where $a_e, b_e$ have been 
defined in eq. (\ref{eq:ffh}), and $f_{1,2}$ are known loop functions \cite{Stockinger:2006zn}. 
Under the assumption that the Higgs-electron coupling is standard, $a_e = 1, b_e = 0$, a rather stringent 
constraint, $b_t < 0.01$, can be realized\cite{Brod:2013cka}. Of course, with different assumptions on 
$a_e, b_e$, or even with additional sources of CP violation, this constraint can become milder or 
evaporate altogether. For example, and as emphasized in ref. \cite{Brod:2013cka}, current Higgs data are 
actually compatible with a Higgs only coupled to third-generation fermions. In this case $b_t$ values of 
O(1) are allowed by the EDM constraints. Furthermore, ref. \cite{Arbey:2014msa} provides another example 
of multi-Higgs scenario, realized in the framework of a CP violating supersymmetric model, in which the
current EDM constraints can be satisfied, in spite of CP violating couplings between the Higgs 
states and the top quark.

It should be noted that, given the smallness of the electron Yukawa coupling, it is unclear whether the 
$a_e, b_e$ couplings will be accessible experimentally in the near future. In order to reconstruct the 
$\tth$ coupling direct probes of the same are necessary, which we will discuss in the next section. 
In this section we focus our attention on the constraints on the $\tth$ coupling that can be derived 
from Higgs rate information collected at the LHC. We will show that these constraints strongly depend 
on the nature of the assumption and one cannot conclusively determine the $\tth$ vertex using signal 
strengths alone.

\subsection{Constraints from measurements of Higgs rates}\label{sec:constr_higgs_rates}

Within the SM, and with Higgs and top masses as measured, there are four main production modes of the 
Higgs at the LHC: gluon fusion (ggF), vector-boson fusion (VBF), Higgs production in association with a 
$W$/$Z$ boson (VH), and Higgs production in association with a $\ttb$ pair. The gluon-fusion production 
mode has the largest cross-section at the LHC, and the dominant contribution to this process comes from 
a top loop. The Higgs decay to two photons has also a contribution due to a top loop, although the 
dominant one comes from a $W$-boson loop. ATLAS and CMS have already put indirect constraints on the 
value of $\at$ in eq. (\ref{eq:ffh}), assuming that there are no other sources contributing to the 
effective couplings $g g \to h$ or $h \to \gamma \gamma$. At $95\%$ confidence level these constraints 
read \cite{CMS-PAS-HIG-14-009,ATLAS-CONF-2014-009}
\bea
\at \in [-1.2,-0.6]\cup[0.6,1.3]&&\qquad 
{\rm ATLAS}\; \nonumber\\
\label{indconstr}
\at \in [0.6,1.2]\qquad\qquad\qquad\; &&\qquad 
{\rm CMS}\;.\nonumber
\eea
%%%%
In this section we extend this analysis by allowing in the fit both $\at$ and $\bt$ couplings in 
eq. (\ref{eq:ffh}), and by including the recently measured $\tth$ channel signal strengths~\cite{Aad:2014lma,ATLAS-CONF-2014-011,Aad:2015gra,Khachatryan:2014qaa,ATLAS-CONF-2015-006}.
Higgs couplings to massive gauge bosons are defined by
\be
\Lag_{hVV}= g m_W h \left(\kappa_W W^{\mu} W _{\mu} 
+ \frac{\kappa_Z}{2 \cw^2}Z^{\mu}Z_{\mu}\right).
\label{eq:kappaV}
\ee
In the SM and at tree level $\kappa_Z=\kappa_W=\kappa_V=1$. 
As customary, the signal strength measured in a particular channel $i$ at the LHC is defined as
\be
\hat{\mu}_i = \frac{n_{\rm exp}^i}{n_{\rm SM}^i}~,
\label{eq:mui_exp}
\ee
where $n_{\rm exp}^i$ is the number of events observed in the channel $i$ and $n_{\rm SM}^i$
is the expected number of events as predicted in the SM. In order to contrast specific model predictions
with the experimentally derived $\hat \mu_i$ we define (as usual)
\be
\mu_i = \frac{n_{\rm th}^i}{n_{\rm SM}^i} = 
\frac{\Sigma_p \sigma_p \epsilon_p^i}{\Sigma_p \sigma_p^{\rm SM} \epsilon_p^i} 
\times \frac{\mc B_i}{\mc B_i^{\rm SM}}~.
\label{eq:mui}
\ee
Here $n_{\rm th}^i$ corresponds to the expected number of events predicted in the hypothesized model 
under consideration; $\sigma_p$ corresponds to the cross-section in the $p^{\rm th}$ production mode, 
i.e. the cross-section for Higgs production in one of the four production modes listed earlier; 
$\mc B_i$ is the branching ratio of the Higgs in the $i^{\rm th}$ channel; $\epsilon^i_p$ is the 
efficiency of the $p^{\rm th}$ production mode to the selection cuts imposed in the $i^{\rm th}$ channel.
Note that the efficiencies in the numerator and denominator of eq. (\ref{eq:mui}) are taken to be the same. 
This is true at leading order for the gluon fusion process. 

In order to evaluate the signal strength in the $\tth $ production channel,
ATLAS~\cite{Aad:2014lma,ATLAS-CONF-2014-011,Aad:2015gra,ATLAS-CONF-2015-006} and CMS~\cite{Khachatryan:2014qaa} first apply some basic 
selection cuts and then use boosted decision trees (BDT) to further separate signal from background. We 
have checked at parton level that for basic selection cuts the efficiency in the two cases of pure scalar 
vs. pure pseudo-scalar Higgs are not significantly different. However, this may not be the case for BDT. 
We neglect this effect here, we namely assume that BDT analyses will have the same efficiency for a scalar 
and a pseudo-scalar Higgs and set them to be equal.

We next discuss the $\at$ and $\bt$ coupling contributions to Higgs production from gluon fusion 
and Higgs decay to two photons. The ratio of the Higgs decay width to two photons to the SM decay width, 
at next to leading order and neglecting the small contribution from fermions other than the top quark, 
can be written in the form \cite{Djouadi:2005gj,Spira:1995rr}
\bea
\label{eq:hyy}
\Rg[\gamgam] &\simeq&
\frac{|\kappa_W A^a_W(\tau_W) + \frac{4}{3} \, \at 
\left( 1 - \alpha_s / \pi \right) \, A^a_t(\tau_t)|^2 
+ |\frac{4}{3} \, \bt \, A^b_t(\tau_t)|^2}
{| A^a_W(\tau_W) + \frac{4}{3} 
\left( 1 - \alpha_s / \pi \right) A^a_t(\tau_t)|^2} \nn \\ 
&\simeq&1.6 \Big( (\kappa_W -0.21 \; \at)^2 + 0.12 \; \bt^2 \Big)~.
\eea
Here $A^i_j$ denote the loop functions due to the $W$ loop ($A^a_W$), the CP-even top coupling ($A^a_t$) 
and its CP-odd counterpart ($A^b_t$). The analytic expressions for these functions are given in 
app.~\ref{sec:app2}. It should be stressed that, given the measured Higgs and top masses, implying 
$\tau_t = m_h^2/(4 m_t^2) \ll 1$, the top contribution (both scalar and pseudo-scalar) is, to a very 
good approximation, given by its expression in the infinite top-mass limit (see app.~\ref{sec:app2}). 
Correspondingly, $\alpha_s$ corrections are included in this limit. In the same limit, they affect only 
the scalar contributions, whereas the the pseudo-scalar one is untouched.

We relate Higgs production through gluon fusion, normalized to the SM value, to the corresponding 
normalized width of Higgs to two gluons. Keeping only the dominant top contribution again, we may write,\footnote{%
The first equality in eq. (\ref{eq:ggh}), relating the widths and the cross section, is an exact equality 
at LO. Luckily, in the heavy top-mass limit ($m_h^2 \ll 4 m_t^2$) and because we are considering ratios of 
$\sigma(gg \to h)$ and ratios of $\Gamma(h \to gg)$, the equality holds to a very good approximation also 
beyond LO. In particular, for the cross section the higher-order long-distance corrections involving 
(infra-red/collinear) emission are universal in this limit. There only remains a genuine higher-order 
correction which depends specifically on the nature of the $t\bar t h$ coupling. A large part of this 
finite regular correction cancels when considering the ratios. This explains the rather small 
$\alpha_s$ correction that we give in eq.~(\ref{eq:ggh}). For a thorough analysis see 
\cite{Djouadi:2005gj,Spira:1995rr}.}

%in the same approximations as eq. (\ref{eq:hyy})
\be
\label{eq:ggh}
\sigh[gg\to h] ~\simeq~ \Rg[gg] ~ \simeq~
\at^2+ \bt^2 
\frac{|A^b_t(\tau_t)|^2}{|A^a_t(\tau_t)|^2}  \left( 1 + \frac{1}{2}\frac{\alpha_s}{\pi} \right)
~\simeq~ \at^2 + 2.29\ \bt^2~.
\ee
Note that, the indirect effect of the pseudo-scalar contribution in $gg \to h$ (and $h \to gg$) is more than twice the corresponding scalar contribution (with $a_t=b_t$). As we will see, in direct $ t \bar t h$ production it is the scalar that 
contributes the most. 
%The coefficients $\tilde{C}^\gamma_{t}$, $\tilde{C}^g_t$, $C^\gamma_{t}$ and $C^g_t$ in eqs. 
%(\ref{eq:hyy}) and (\ref{eq:ggh}) contain the NLO QCD corrections to the LO expressions and are detailed 
%in appendix~\ref{sec:app2}. The use of these NLO factors is, besides the use of the latest data from 
%ATLAS and CMS and the inclusion of $\tth$ production data, the main difference between the fits performed here and those existing in the literature.\footnote{See for example ref.~\cite{Djouadi:2013qya}.}

We now perform a global fit to the Higgs data collected by ATLAS, CMS and Tevatron in order to estimate 
the allowed values of $\at$ and $\bt$. We follow closely the procedures of 
refs.~\cite{Cacciapaglia:2012wb,Djouadi:2013qya,Cheung:2013kla}.

In general, BSM models allow for additional interactions not present in the SM to both the scalar and 
pseudo-scalar components of the Higgs, that may be CP-conserving or not.
Gluon fusion and Higgs to di-photon decays, being loop-induced processes, are sensitive probes of this 
new physics. In this sense unknown heavy physics  not related to the top could contribute to the effective 
operators describing gluon fusion ($h G^{\mu \nu} G_{\mu \nu}, h G^{\mu \nu} \tilde{G}_{\mu \nu}$, where 
$G_{\mu \nu}$ is the gluon field strength and $\tilde{G}_{\mu \nu}$ its dual) and decays into photons ($h 
F^{\mu \nu} F_{\mu \nu}, h F^{\mu \nu} \tilde{F}_{\mu \nu}$, $F_{\mu \nu}$ and $\tilde{F}_{\mu \nu}$ 
denoting again the electromagnetic field strength and its dual, respectively). In order to account for 
these additional BSM effects, following ref.~\cite{Cacciapaglia:2012wb}, we introduce four extra 
parameters $\kappa_{gg}$, $\tilde{\kappa}_{gg}$, $\kappa_{\gamma\gamma}$ and 
$\tilde{\kappa}_{\gamma\gamma}$ so that eqs. (\ref{eq:hyy}) and (\ref{eq:ggh}) are modified as follows
\begin{eqnarray}
\Gamma_{\gamma \gamma}/\Gamma^{\rm SM}_{\gamma \gamma} &\simeq&
1.6 \Big( (\kappa_W - 0.21 \; (\at + \kappa_{\gamma \gamma}) )^2 
+ 0.12 \; (\bt +\tilde \kappa_{\gamma \gamma})^2 \Big)~, \nn \\
\Gamma_{g g}/\Gamma^{\rm SM}_{g g} &\simeq&
(\at + \kappa_{gg})^2 + 2.29\ (\bt + \tilde \kappa_{gg})^2~.
\label{eq:kappaVV}
\end{eqnarray}
Note namely that the couplings $\kappa_{gg}$, $\tilde{\kappa}_{gg}$, $\kappa_{\gamma\gamma}$ and 
$\tilde{\kappa}_{\gamma\gamma}$ are normalized so that, in these observables, they shift $a_t$ and $b_t$ 
with a relative factor of unity.
% THESE EQS GO IN THE APPENDIX
%\begin{eqnarray}
%\label{eq:dec}
%\Gamma_{\gamma \gamma} &=& 
%\frac{G_F \alpha^2 m_H^3}{128 \sqrt{2} \pi^3} 
%\bigg\{ \left|  
%\kappa_W \, A_W^{a} (\tau_W) + \frac{4}{3} (a_t+\kappa_{\gamma \gamma}) \, C^\gamma_t \, 
%A_t^{a} (\tau_t)
%\right|^2 \nonumber \\
%&+& 
%\left| 
%\frac{4}{3} ( \bt + \tilde{\kappa}_{\gamma\gamma} ) \, \tilde{C}^{\gamma}_{t} \, A^b_t(\tau_t)
%\right|^2
%\bigg\}~, \\
%\Gamma_{g g} &=& 
%\frac{G_F \alpha_s^2 m_H^3}{16 \sqrt{2} \pi^3}
%\bigg\{ \left| 
%\frac{1}{2} (a_t+\kappa_{gg}) \, C^g_t \, A_t^{a} (\tau_t) 
%\right|^2 \nonumber \\
%&+&
%\left|
%\frac{1}{2} (\bt+\tilde{\kappa}_{gg}) \, \tilde{C}^g_t \, A_t^{b} (\tau_t)
%\right|^2 \bigg\}
%\label{eq:kappaVV}~.
%\end{eqnarray}
%

The fit to $\mu_i$ is performed by minimizing the $\chi^2$ function defined as 
\be
\label{eq:chi2}
\chi^2 =\sum\limits_{i}\left(\frac{\mu_i -\hat{\mu}_i}{\hat \sigma_i}\right)^2,
\ee
where $\hat \mu_i$ are the experimental measurements and $\hat \sigma_i$ their uncertainties.
We take into account the possibility of asymmetric errors by using the prescription 
of ref.~\cite{Banerjee:2012xc}.
Namely, whenever errors are quoted as $(\hat{\mu}_i)^{+y}_{-z}$, we take $ \hat \sigma_i = y$
if $(\mu_i -\hat{\mu}_i) >0$, and $\hat \sigma_i = z$ if $(\mu_i -\hat{\mu}_i) <0$~\cite{Banerjee:2012xc}.
In some of the measured channels, the experimental collaborations have provided information on the 
correlation between different production modes. In this case we modify the $\chi^2$ function to include 
these correlations as follows
\be
\label{eq:chi2_corr}
\chi^2(i,j)=\frac{1}{1-\rho^2}\left[
\left(\frac{\mu_i -\hat{\mu}_i}{\hat \sigma_i}\right)^2 
+ \left(\frac{\mu_j -\hat{\mu}_j}{\hat \sigma_j}\right)^2
-2\rho\left(\frac{\mu_i -\hat{\mu}_i}{\hat \sigma_i}\right)
\left(\frac{\mu_j -\hat{\mu}_j}{\hat \sigma_j}\right)
\right]
\ee
where $\rho$ is the correlation coefficient and $i$ and $j$ correspond to different Higgs production modes.
The data used in the fits are detailed in app.~\ref{sec:app1}. 

%%%%%%%%%%%%%%%%%%%%%%%%%%%%%%%%%%%%%%%%%%%%%%%%%%%%%%%%%%%%%%%%%%%%%%%%
\subsubsection{Results}
%%%%%%%%%%%%%%%%%%%%%%%%%%%%%%%%%%%%%%%%%%%%%%%%%%%%%%%%%%%%%%%%%%%%%%%%
\begin{figure}
\centering
\includegraphics[scale=0.6]{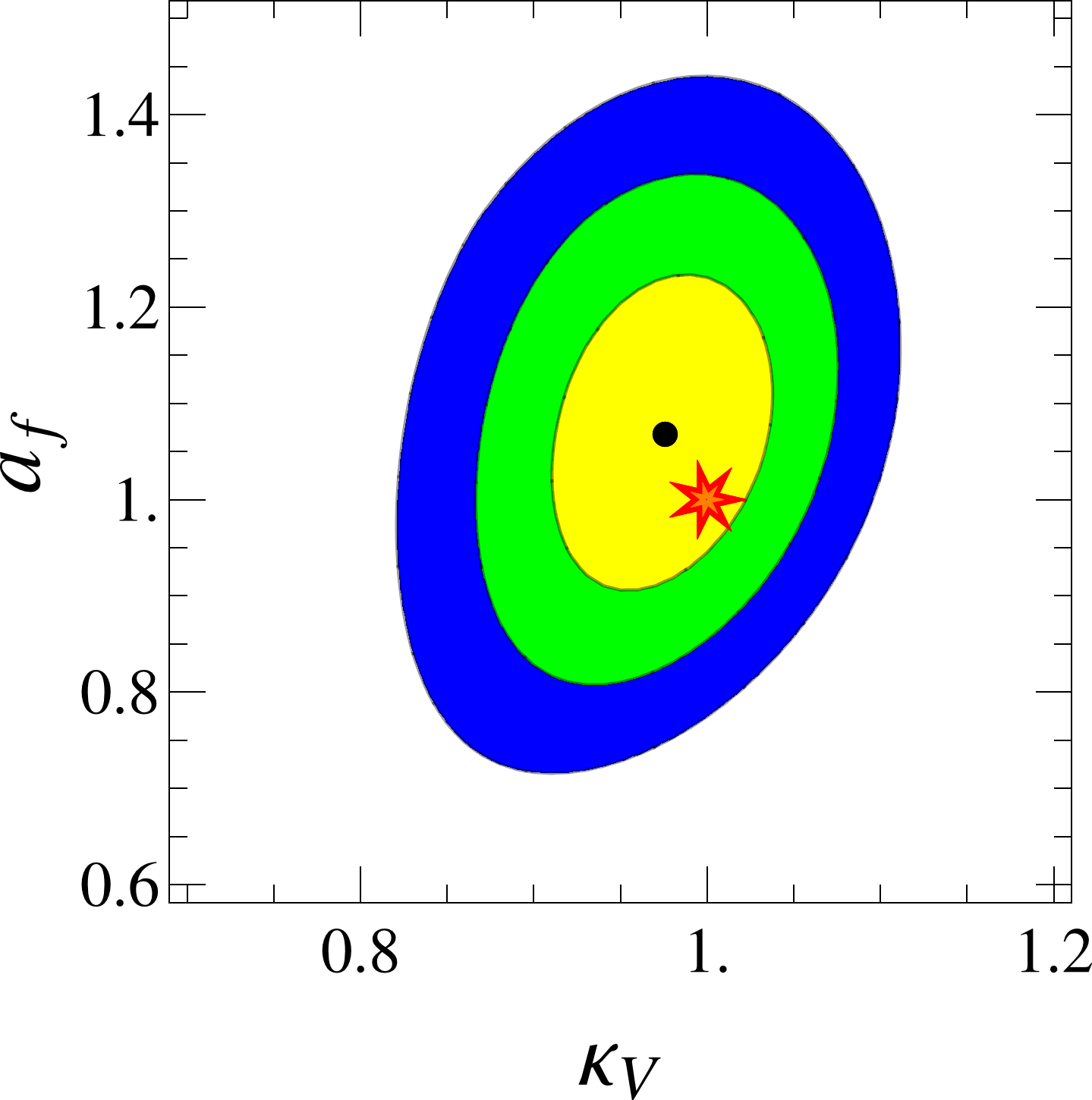}
\caption{ 
Fit results for $a_f$ vs. $\kappa_V$. The black dot indicates the best-fit value. The yellow (white), 
green (medium grey) and blue (dark grey) areas represent the $68\%$, $95\%$ and $99.7\%$ confidence-level 
regions, respectively. The red star shows the SM point $(\kappa_V,a_f)=(1,1)$.
}
\label{fig:fit-sm}
\end{figure}

We first perform a fit to the SM couplings $a_f$ and $\kappa_V$, while setting all other couplings to zero.
The results of this fit are displayed in fig. \ref{fig:fit-sm}. Here we only show the contours for positive 
values of $a_f$ and $\kappa_V$. An excess seen initially in the $h\to \gamma\gamma$ channel
(excess which is now reduced in ATLAS data and absent in CMS data) pointed to negative values of $a_f$, 
which would have had serious consequences on unitarity \cite{Bhattacharyya:2012tj,Choudhury:2012tk}. In 
the figure, the black dot at (0.97,\,1.06) indicates the best-fit value, while the yellow, green and blue regions correspond 
to the $68\%$, $95\%$ and $99.7\%$ confidence level regions, respectively. The SM value of 
$(\kappa_V,a_f)=(1,1)$ is indicated by a red star. Analyses performed by CMS~\cite{CMS-PAS-HIG-13-005} 
find a best-fit value at slightly smaller values of $\kappa_V$, while fits performed by 
ATLAS~\cite{ATLAS-CONF-2013-034} indicate larger values of $\kappa_V$. Since we have used both sets of 
data, we arrive at a middle point, in very good agreement with the SM expectation. We found good agreement 
with the fits of ATLAS and CMS when we use only their respective data sets. 
\begin{figure}[b!]
\centering
\includegraphics[scale=0.56]{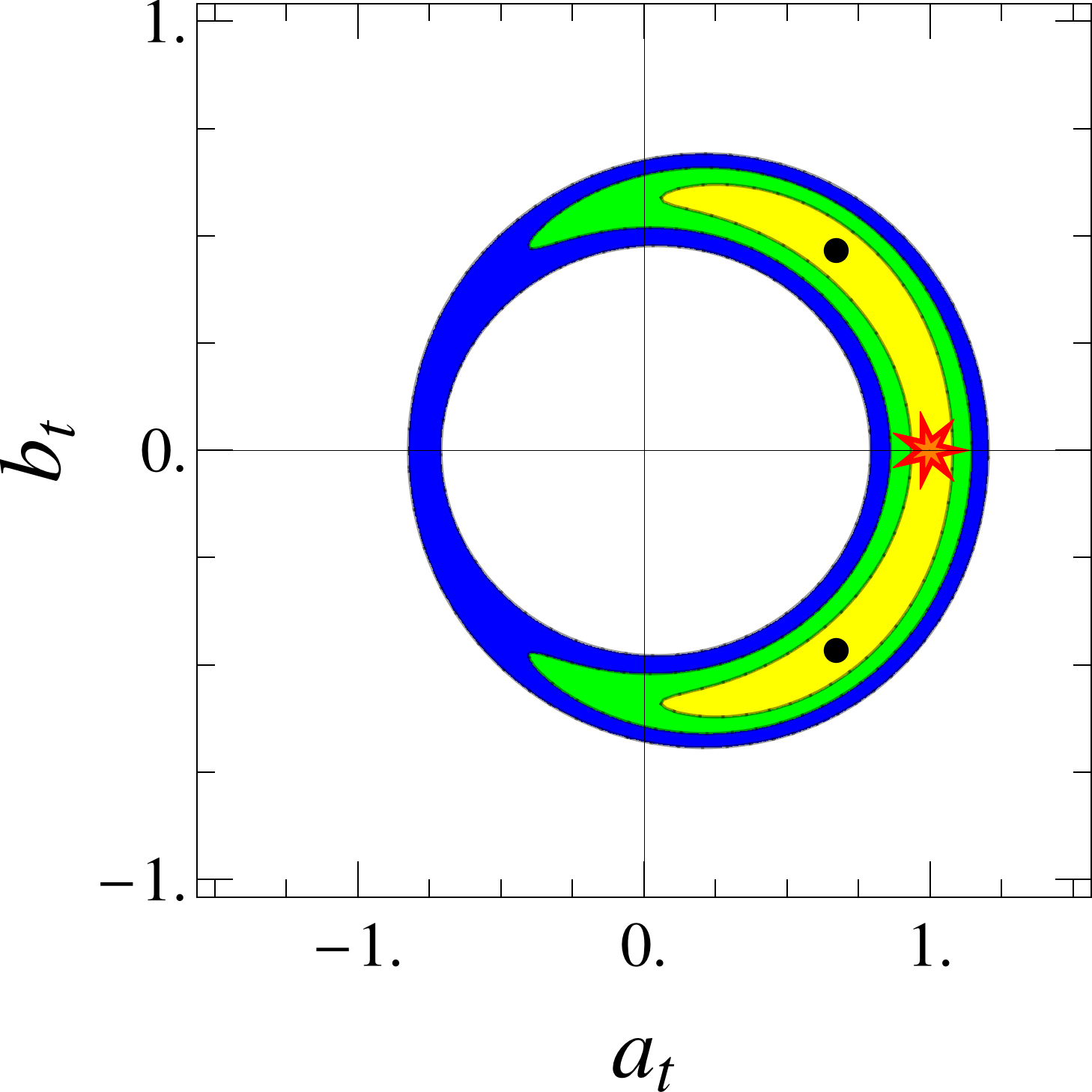} \quad
\caption{
Fit results for $\at$ vs. $\bt$. Black dots indicate the best-fit values. The yellow (white), 
green (medium grey) and blue (dark grey) areas represent the $68\%$, $95\%$ and $99.7\%$ confidence-level 
regions, respectively. The red star shows the SM point $(a_t,\bt)=(1,0)$. The fit to $\at, \bt$ is 
performed while keeping all other parameters fixed to their SM values, i.e. $\kappa_V = 1, a_f = 1$ and 
$b_f = 0$ for any $f \neq t$.
\label{fig:fit-cp}
}
\end{figure}

We next perform a fit to the parameters $\at$ and $\bt$ -- the CP-even and CP-odd Higgs-top quark 
couplings. All other parameters are fixed to their SM values, i.e. $\kappa_V=1$, $a_f=1$
and $b_f=0$ for any $f \neq t$. The results of this fit are shown in fig. \ref{fig:fit-cp}.
Similar analyses have also been performed in refs.~\cite{Djouadi:2013qya,Cheung:2013kla,Nishiwaki:2013cma}.
We find two best-fit values, $(\at,\bt)=(0.67,0.46)$ and $(\at,\bt)=(0.67,-0.46)$, which are symmetric 
about the $\bt$ axis as expected. Remarkably, significant departures from the SM expectation are still 
possible for the CP-odd coupling. The shape of the $68\%$, $95\%$ and $99.7\%$ confidence level regions 
in fig.~\ref{fig:fit-cp} can be easily understood by looking at eqs. (\ref{eq:hyy}) and (\ref{eq:ggh}).
In $gg \rightarrow h$ the $\at$ and $\bt$ coefficients enter quadratically, weighed by the loop functions 
$A_t^a$ and $A_t^b$, respectively. Therefore, while $gg \rightarrow h$ production is useful to constrain 
the overall $\at^2$ and $\bt^2$ magnitudes, alone it is unable to distinguish between scalar and 
pseudo-scalar effects. Inclusion of the $h \rightarrow \gamma \gamma$ decay channel substantially 
improves the discriminating power. 
The important point is that, in this decay channel, the scalar-coupling contribution, contrary to the 
pseudo-scalar one, interferes with the $W$ contribution. In particular, for $\at >0$, as in the SM, this 
interference is destructive. On the other hand, for $\at$ negative, the branching ratio gets enhanced with 
respect to the SM one by both the scalar and the pseudo-scalar contributions, thus making 
$\Gamma(h \rightarrow \gamma \gamma)$ too large. This is the reason why $\at<0$ is less favoured than 
$\at>0$ in fig. \ref{fig:fit-cp}. 
Specifically, $\at=0$ does not fit the data either because in this case the $W$ loop is too large and 
cannot obviously be compensated by the $\bt$ contribution, irrespective of the value of $\bt$. 

%\begin{figure}[h!]
%\centering
%\includegraphics[scale=0.56]{c7-figs/atbt_gaga.png}
%\includegraphics[scale=0.56]{c7-figs/atbt_gaga_proj.png} 
%\caption{
%Fits to $\at$ vs. $\bt$ in a scenario with $\tilde{\kappa}_{gg}=\tilde{\kappa}_{\gamma\gamma}=-1$ and 
%$\kappa_{gg}=\kappa_{\gamma\gamma}=0$. Black dots indicate the best-fit values. The yellow (white), 
%green (medium grey) and blue (dark grey) areas represent the $68\%$, $95\%$ and $99.7\%$ confidence-level 
%regions, respectively. The red star shows the SM point $(a_t,\bt)=(1,0)$.
%The plot to the right shows the same fit as the plot to the left, but with pseudo-data, with namely
%all the signal strengths set to the SM values and the uncertainties reduced by half.
%\label{fig:fit-cp2}
%}
%\end{figure}

Next we look at the effect of the parameters $\kappa_{gg}$, $\kappa_{\gamma\gamma}$, 
$\tilde{\kappa}_{gg}$ and $\tilde{\kappa}_{\gamma\gamma}$, introduced in eqs.~(\ref{eq:kappaVV}).
In particular, we would like to investigate their impact on the value of $b_t$. By inspection 
of eqs. (\ref{eq:kappaVV}), it is clear that an arbitrary (common) value for $\tilde{\kappa}_{gg}$ and 
$\tilde{\kappa}_{\gamma\gamma}$ can always be compensated by $b_t$. Therefore, a simultaneous fit
of Higgs-rate data to $\tilde{\kappa}_{gg}$, $\tilde{\kappa}_{\gamma\gamma}$ and $b_t$ would result in
the flat direction $|b_t| = - |\tilde{\kappa}_{gg}| = -|\tilde{\kappa}_{\gamma\gamma}|$, with
$|b_t|$ arbitrary.
We quote, as an example, a fit where we set $\tilde{\kappa}_{gg}=\tilde{\kappa}_{\gamma\gamma}=-1$, 
and $\kappa_{gg}=\kappa_{\gamma\gamma}=0$. We find the best-fit point $(\at=0.67,\bt=1.46)$ 
and that the various confidence level contours have shifted by $+1$ in the $b_t$ direction, 
allowing for correspondingly larger values of $\bt$ than the fit of fig. \ref{fig:fit-cp}. 
We would expect a second best-fit point at 
$(\at=0.67,\bt=0.54)$, according to the discussion in the previous paragraph, and as displayed in fig.
\ref{fig:fit-cp}. We actually find that the $\chi^2$ value of this second solution is not exactly equal 
to the $\chi^2$ at the best-fit point, although the relative difference is puny, $2 \times 10^{-5}$. 
Exact degeneracy is lifted by the tagged $\tth$ data in table \ref{c7:t6}, to which the 
$\pm (b_t + \tilde \kappa_{gg, \gamma \gamma})$ symmetry does not apply, at variance with the rest of
the data. This example demonstrates that $\tth$ data would {\em in principle} be able to resolve the
degeneracy in the $b_t$ solutions, but their discriminating power is limited by their small statistical
weight as compared with the rest of Higgs-rate data.
%As an example, in fig. \ref{fig:fit-cp2} we show the results of the fits when we set 
%$\tilde{\kappa}_{gg}=\tilde{\kappa}_{\gamma\gamma}=-1$, and $\kappa_{gg}=\kappa_{\gamma\gamma}=0$. 
%Note that the assumption $\tilde{\kappa}_{gg}=\tilde{\kappa}_{\gamma\gamma}=-1$ excludes the SM point
%by definition, as displayed in fig. \ref{fig:fit-cp2}.
%We find the best-fit point $(\at=0.67,\bt=1.46)$ and that the various confidence level contours have 
%shifted upwards, allowing for much larger values of $\bt$. The right plot of the same figure shows the 
%same fit, but using pseudo-data with all the signal strengths set to be exactly equal to the SM value 
%and with uncertainties reduced by half.\footnote{In addition, we also include a signal 
%strength related to the $h\to Z \gamma$ channel, with uncertainties to be the same as the $h\to \gamgam$ 
%channel.} This has been done in anticipation of the LHC data at 14 TeV. We see that even in this case 
%large values of $\bt$ are still allowed.

Altogether, this example is meant to show the inherent limitation of using indirect effects to probe the 
$b_t$ interaction.
As a matter of fact, in spite of using a very minimal set of parameters, data does not rule out a non-zero 
$\bt$. Furthermore, on introducing additional sources of pseudo-scalar interactions, even larger values of
$\bt$ can be accommodated. Finally, since signal strengths are CP-even quantities (and therefore not 
linear in $\bt$), they do not provide information on the sign of $\bt$.
All such ambiguities in the determination of $\at$ and $\bt$ could only be resolved with more direct 
probes, as discussed in the remainder of this work.

%%%%%%%%%%%%%%%%%%%%%%%%%%%%%%%%%%%%%%%%%%%%%%%%%%%%%%%%%%%%
% % % % % % % % % % %: Kinematics of tth % % % % % % % % % %
%%%%%%%%%%%%%%%%%%%%%%%%%%%%%%%%%%%%%%%%%%%%%%%%%%%%%%%%%%%%
\section{Associated production of the Higgs with a $\ttb$ pair} \label{sec:tth_obs}

\subsection{Kinematics of $\tth$ production: scalar- vs. pseudo-scalar-Higgs cases}

Of the four production modes (ggF, VBF, VH, $\tth$, with $V=W^\pm, Z$) of the Higgs at the LHC, 
$\tth$ production has the smallest cross-section. Search strategies for the $\tth$ process at the LHC 
have been studied in various Higgs 
decay modes~\cite{ATLAS_TDR,Ball:2007zza}: $b\bar{b}$ \cite{Plehn:2009rk,Artoisenet:2013vfa}, 
$\tau^{+}\tau^{-}$ \cite{Buckley:2013auc} and $W^{+} W^{-}$ \cite{Maltoni:2002jr,Curtin:2013zua,%
Agrawal:2013owa}.
The complicated final state of the process, with 
the top quark decaying to a bottom quark and a $W$ boson, which in turn may decay either hadronically 
or leptonically, as well as the large backgrounds to the process make this a difficult channel to study at the 
LHC. Note, on the other hand, that $\tth$ production can be studied very precisely at a future linear
collider such as the ILC \cite{Djouadi:2007ik}. Sufficiently high rates for this process are possible 
at such colliders \cite{Djouadi:1992gp,Djouadi:1991tk,Grzadkowski:1999ye,Dawson:1997im,Dittmaier:1998dz,%
You:2003zq,Belanger:2003nm} and can therefore be used to extract CP information \cite{BarShalom:1995jb,%
Atwood:2000tu,Gunion:1996vv,Bhupal-Dev:2007is,Godbole:2011hw,Godbole:2007uz,Huang:2001ns,%
Ananthanarayan:2013cia,Ananthanarayan:2014eea} by exploiting angular correlations and/or polarization 
of the top pair.

As noted in the previous section, studying $\tth$ production at the LHC,  though challenging, is a 
necessary undertaking; among the other reasons in order to unambiguously determine the parity of the 
Higgs coupling to the top quark, and  to reveal potential CP-violating effects in the Higgs-top coupling. In this section we wish to point out the major differences in the kinematics of the top and Higgs that a 
scalar- vs. a pseudo-scalar Higgs entails for $\tth$ production at the LHC. 
This has been discussed in the literature in quite some detail. See for example refs. \cite{Demartin:2014fia,Frederix:2011zi,Accomando:2006ga,Godbole:2004xe,Gunion:1996xu,Gunion:1996vv,He:2014xla,Khatibi:2014bsa} and 
references therein for studies of the CP nature of the $\tth$ vertex at the LHC. Many of these employ 
optimal observables \cite{Gunion:1996vv,Atwood:1991ka} or the modern incarnation of the technique, the multivariate analysis.
%Many of these employ the optimal variable\cite{Atwood:1991ka,Gunion:1996vv} or multivariate analyses.
The aim of the present work at large, is to search for and explore {\it lab-frame observables} able to probe the nature of the $\tth$ interactions at the LHC, in 
spite of the hadronic environment. Our analyses are performed on $14$~TeV LHC collisions at the parton 
level, simulated thanks to the {\tt MadGraph} package \cite{Alwall:2014hca}. In view of the exploratory 
nature of this study, we do not consider the effects of NLO corrections, backgrounds, hadronization, 
initial and final state radiation or detector effects. All the necessary qualifications will be made 
as our observables are discussed.

\begin{figure}
\centering
\includegraphics[scale=0.2]{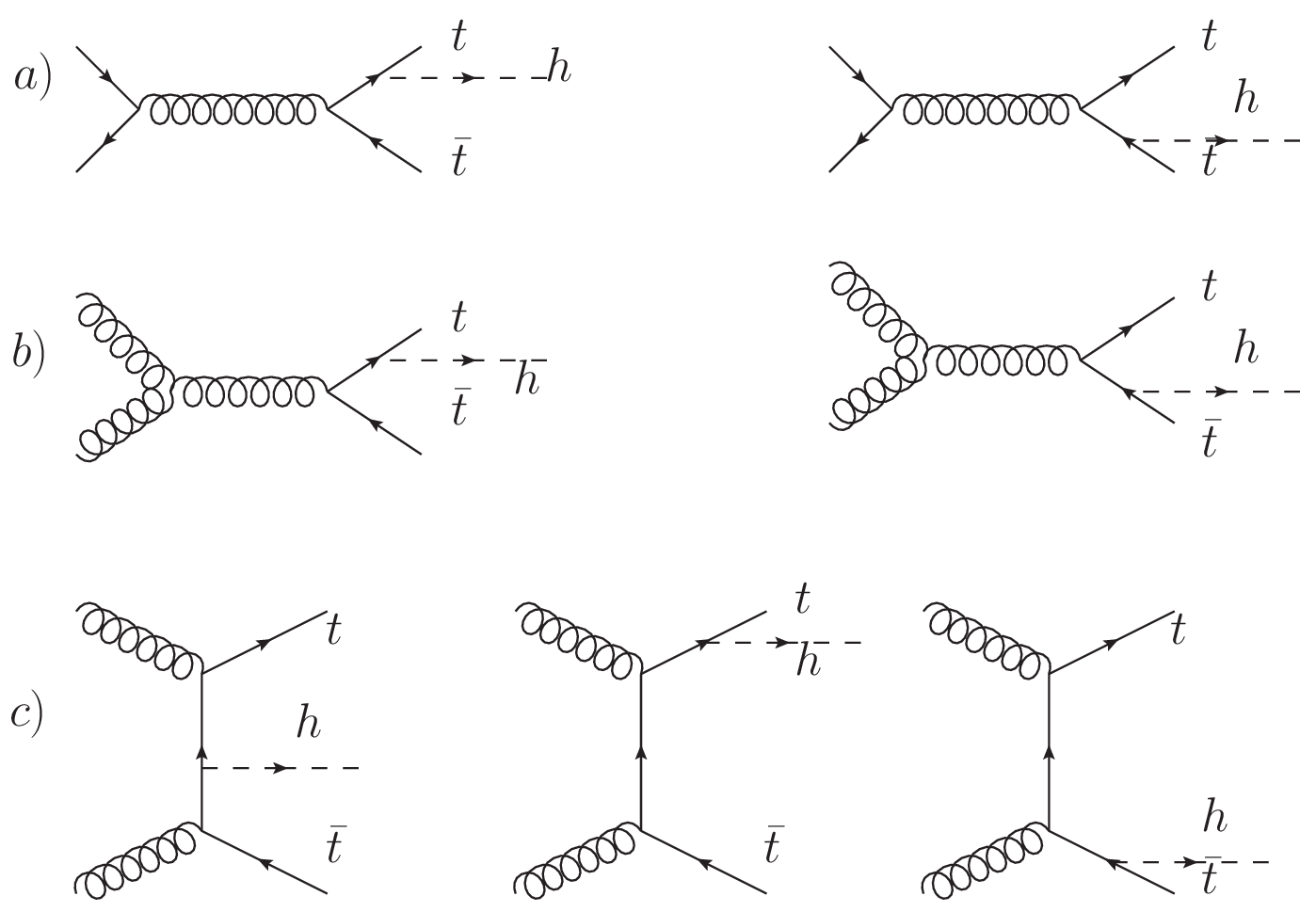}\
\caption{Feynman diagrams for Higgs production in association with a $\ttb$ pair at the LHC. Diagrams 
where the production is mediated by a $Z$ boson or a photon have been omitted. Three more diagrams can 
be realized by exchanging the two gluon lines in the last row labelled (c).
\label{fig:tth-feyn}
}
\end{figure}

As a first step, let us try to understand the kinematics of $\tth$ production without considering the 
decays of the Higgs or the $\ttb$ quarks. The relevant Feynman diagrams are shown in 
fig. \ref{fig:tth-feyn}. We have grouped the diagrams into three categories: quark-initiated, 
gluon-initiated $s$-channel and gluon-initiated $t$-channel. Diagrams where the production is mediated by a 
$Z$ boson or a photon have been omitted. Three more diagrams can be realized by exchanging the two 
gluon lines in the last row labelled (c).

The first distribution we consider is the production cross-section near threshold. It has been pointed 
out that the threshold behaviour of the cross-section for a scalar {\it vs.} a pseudo-scalar Higgs is 
very different at an $e^+ e^-$ collider \cite{Godbole:2011hw,Godbole:2007uz,Bhupal-Dev:2007is}. More
specifically, the {\em rate} of increase of the cross-section with the centre of mass energy of the 
collision is suppressed in the case of the pseudo-scalar Higgs coupling by a factor of $\rho$, where 
$\rho = (\sqrt{s} - 2 m_t - m_h)/\sqrt{s}\ $ parametrizes the proximity to the production threshold. 
This factor can be easily understood from arguments of parity and angular-momentum conservation 
\cite{Bhupal-Dev:2007is}. Close to the energy threshold, the simultaneous demand of angular momentum 
and parity conservation implies that for a scalar the total angular momentum of the $\tth$ system will 
be zero, while for a pseudo-scalar it will be one. Since the process is mediated through $s$-channel 
production, the pseudo-scalar production will be suppressed near threshold.
Note that the total cross-section and not just the behaviour near the threshold is different for a scalar and a pseudo-scalar for the same Yukawa coupling strength.
\begin{figure}
\centering
\includegraphics[scale=0.18]{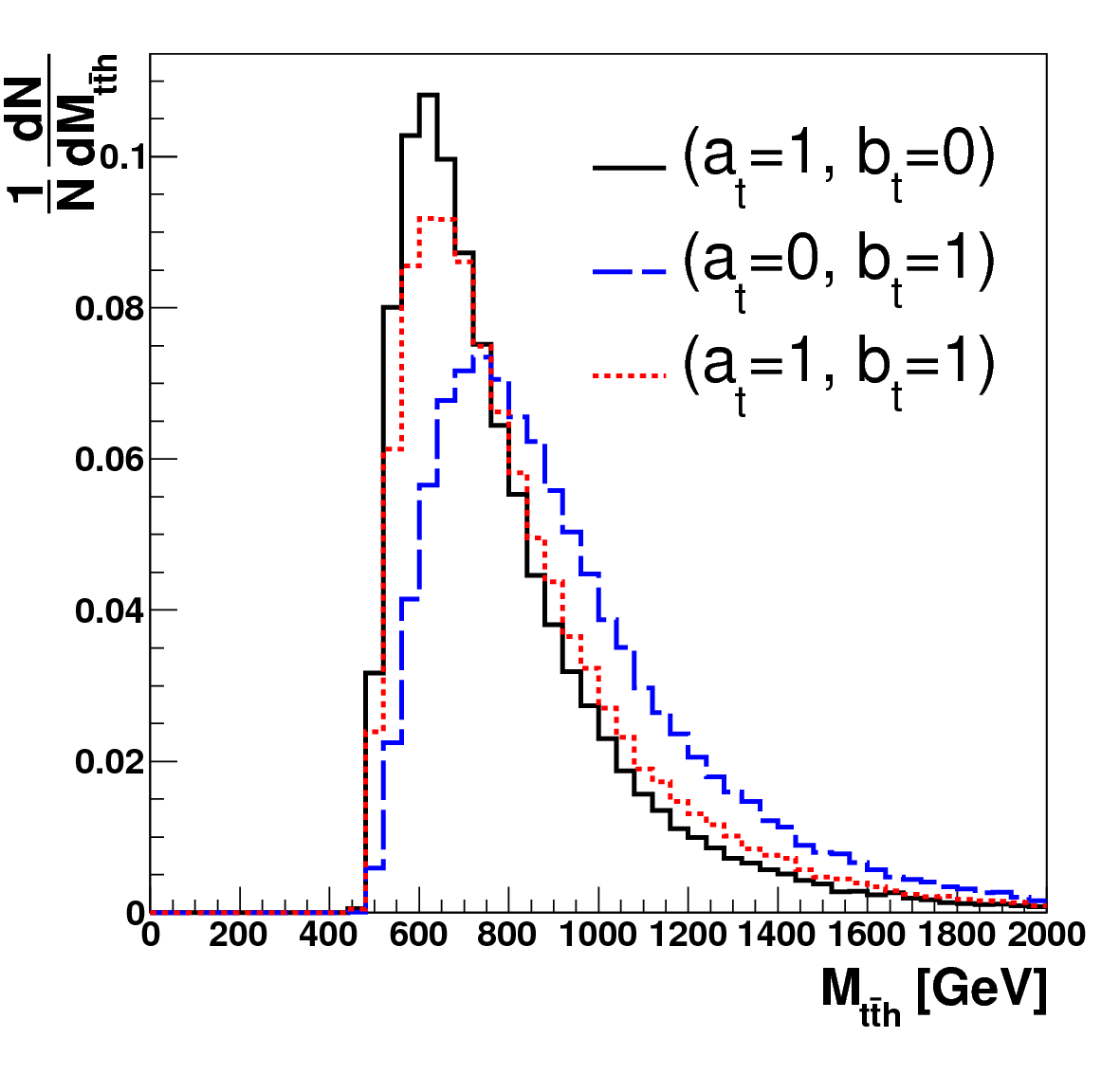}
\includegraphics[scale=0.18]{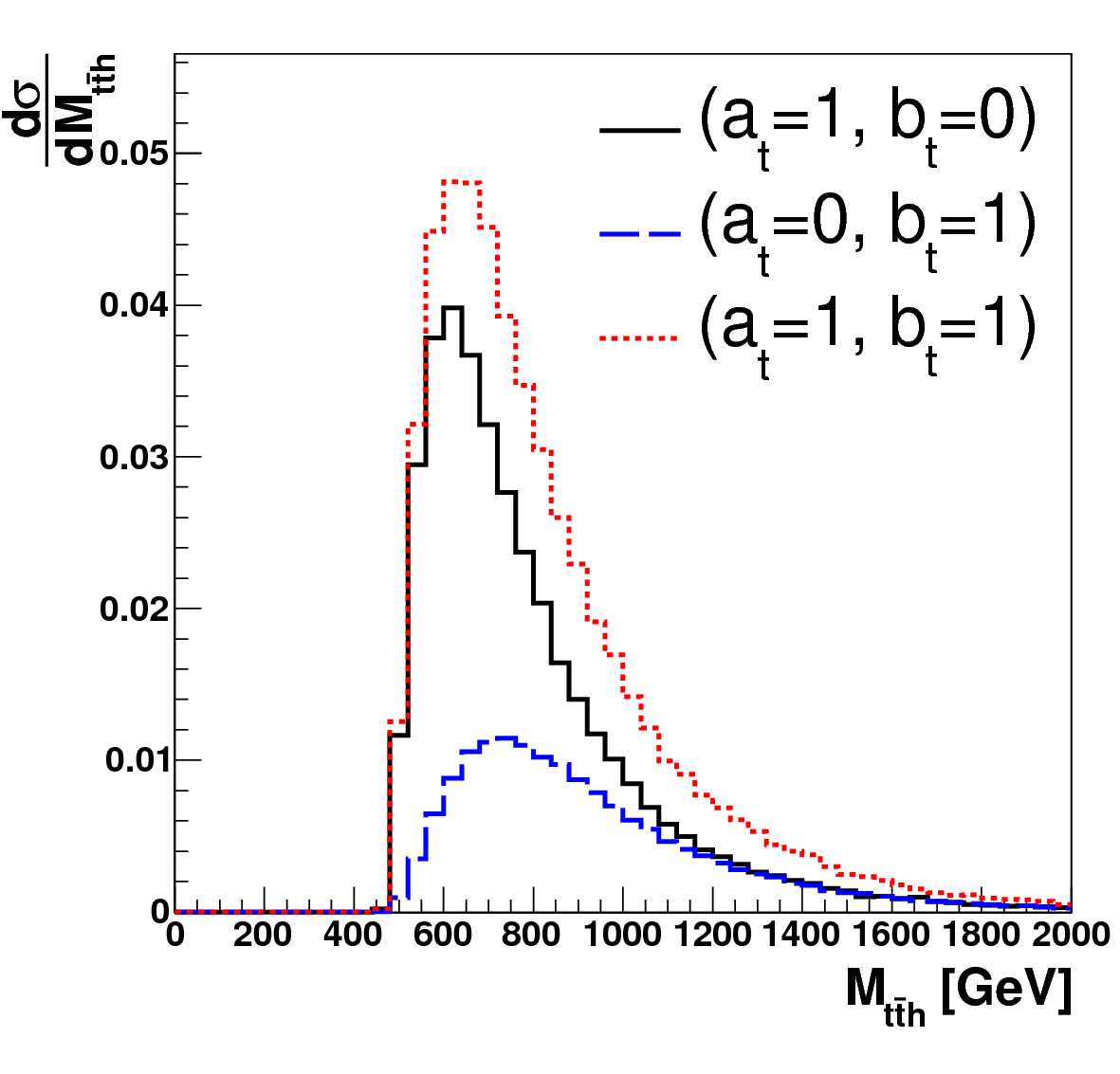}
\caption{
{\em (Left panel)} The invariant-mass distribution of the $\tth$ system, normalized to unity.
{\em (Right panel)} The differential cross-section with respect to the $\tth$ invariant mass. 
In either panel, the SM  distribution ($\at=1,\bt=0$) is shown with a solid black line, the 
pseudo-scalar case ($\at=0,\bt=1$) with a blue dashed line, and the CP-violating case ($\at=1,\bt=1$) 
with a dotted red line.
\label{fig:mtth}
}
\end{figure}

At the LHC, several competing production mechanisms are at work, and it is non-trivial that a 
similar difference in the threshold rise be also visible. 
Indeed the same behaviour as in the $e^+ e^-$ case is observed in the 
quark-initiated process of $pp$ collisions, which is a spin-1, $s$-channel process, but this contribution 
is negligible at the LHC. The dominant $gg$-initiated process, has contributions from 
both $s$-channel and $t$-channel diagrams as shown in fig. \ref{fig:tth-feyn}. While for pseudo-scalar 
production the $s$-channel displays a similar suppression by $\rho$ near threshold, the $t$-channel does 
not. We find however that the cross-section near the production threshold in the $t$-channel displays a 
suppression by a factor proportional to $(m_h/m_t)^4$.
As a result, the production cross-section near threshold does show interesting behaviour.

In the left panel of fig. \ref{fig:mtth} we show the normalized invariant mass distributions of the 
$\tth$ system for the pseudo-scalar $(\at=0,\bt=1)$, the scalar $(\at=1,\bt=0)$ case and the CP-violating case.

We see that the rate of increase of the cross-section with the invariant mass of the $\tth$ system is 
much more rapid for the scalar than for the pseudo-scalar case. This is an important distinguishing 
feature and could be used to probe the nature of the Higgs-top quark coupling. The right panel of 
fig. \ref{fig:mtth} shows the same distributions, but normalized to the total cross-section (i.e. 
$d\sigma/dM_{\tth}$). We observe, as expected, that for the same coupling magnitude, the cross-section 
for the pseudo-scalar case is suppressed with respect to the scalar case.

While the invariant mass distribution is a useful observable to probe the nature of the Higgs-top 
couplings, its measurement is not straightforward. In fact, it requires complete knowledge of the 
top and Higgs momenta, whose reconstruction is challenged by uncertainties on jet energies and, in 
particular, by missing energy, in decay channels including neutrinos.

We note incidentally that, rather than trying to extract the full distribution itself, it might be easier to consider ratios of cross-sections in two $M_{\tth}$ intervals.

\begin{figure}
\centering
\includegraphics[scale=0.18]{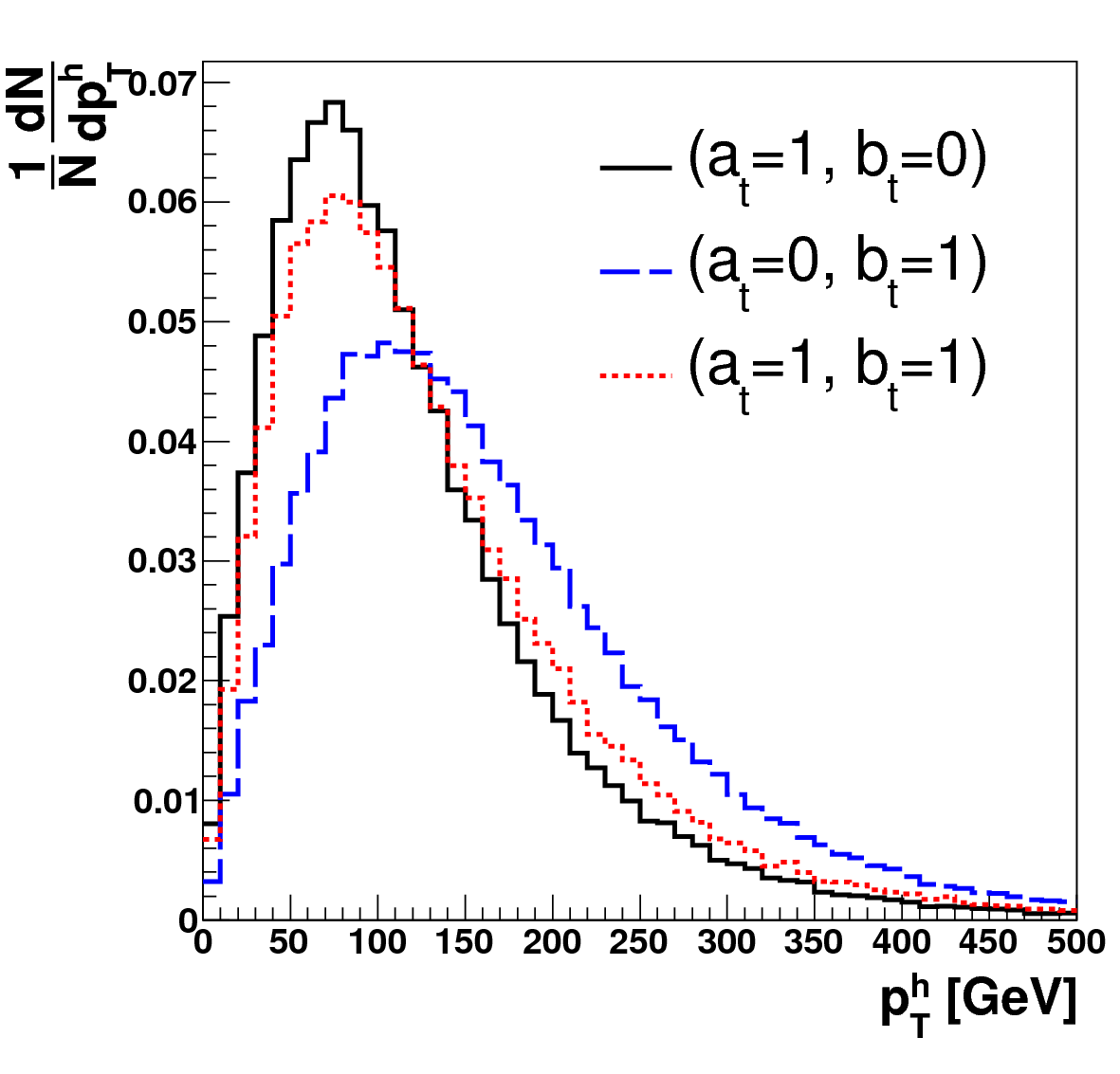}
\includegraphics[scale=0.18]{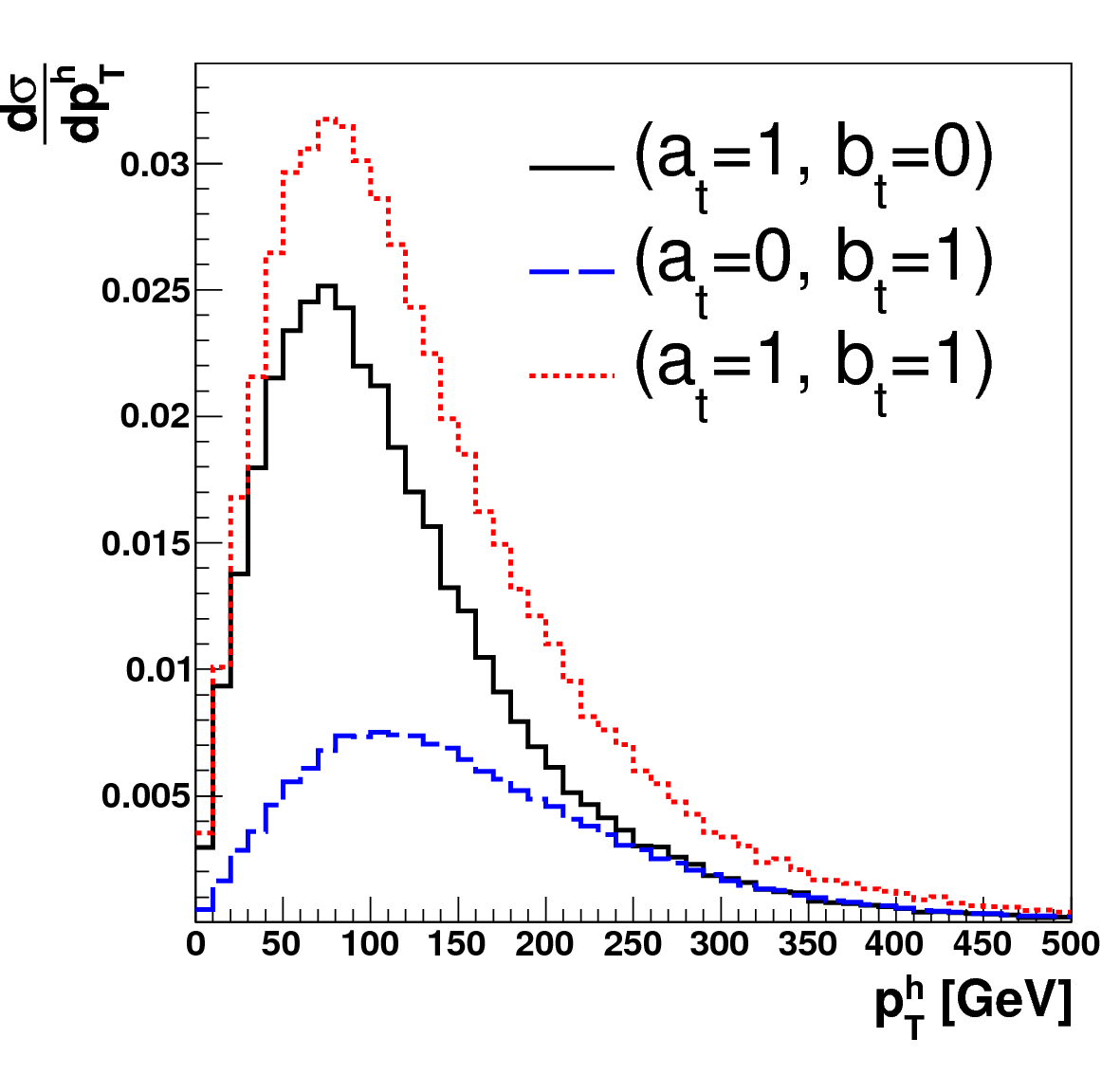}
\caption{
{\em (Left panel)} The distribution of the transverse momentum of the Higgs ($\pt^{h}$), normalized to 
unity. {\em (Right panel)} The differential cross-section with respect to the transverse momentum of 
the Higgs ($\pt^{h}$). In either panel, the SM  distribution ($\at=1,\bt=0$) is shown with a solid black
line, the pseudo-scalar case ($\at=0,\bt=1$) with a blue dashed line, and the CP violating case 
($\at=1,\bt=1$) with a dotted red line.
\label{fig:pth}
}
\end{figure}
The complications mentioned above motivate us to look for alternatives to the invariant-mass 
distribution $M_{\tth}$.
One first possibility, that has also been considered in refs.~\cite{Frederix:2011zi,Demartin:2014fia}, 
is the transverse momentum of the Higgs. Its distributions are shown in fig. \ref{fig:pth},
with normalizations analogous to fig. \ref{fig:mtth}. As a general feature, we note that the transverse 
momentum of the Higgs ($\pt^{h}$) displays a behaviour akin to the invariant-mass distribution $M_{\tth}$. 
Noteworthy is the fact that $\pt^{h}$ is pushed to larger values in the pseudo-scalar case ($\at=0,\bt=1$)
in comparison to the SM distribution ($\at=1,\bt=0$).

\begin{figure}
\centering
\includegraphics[scale=0.2]{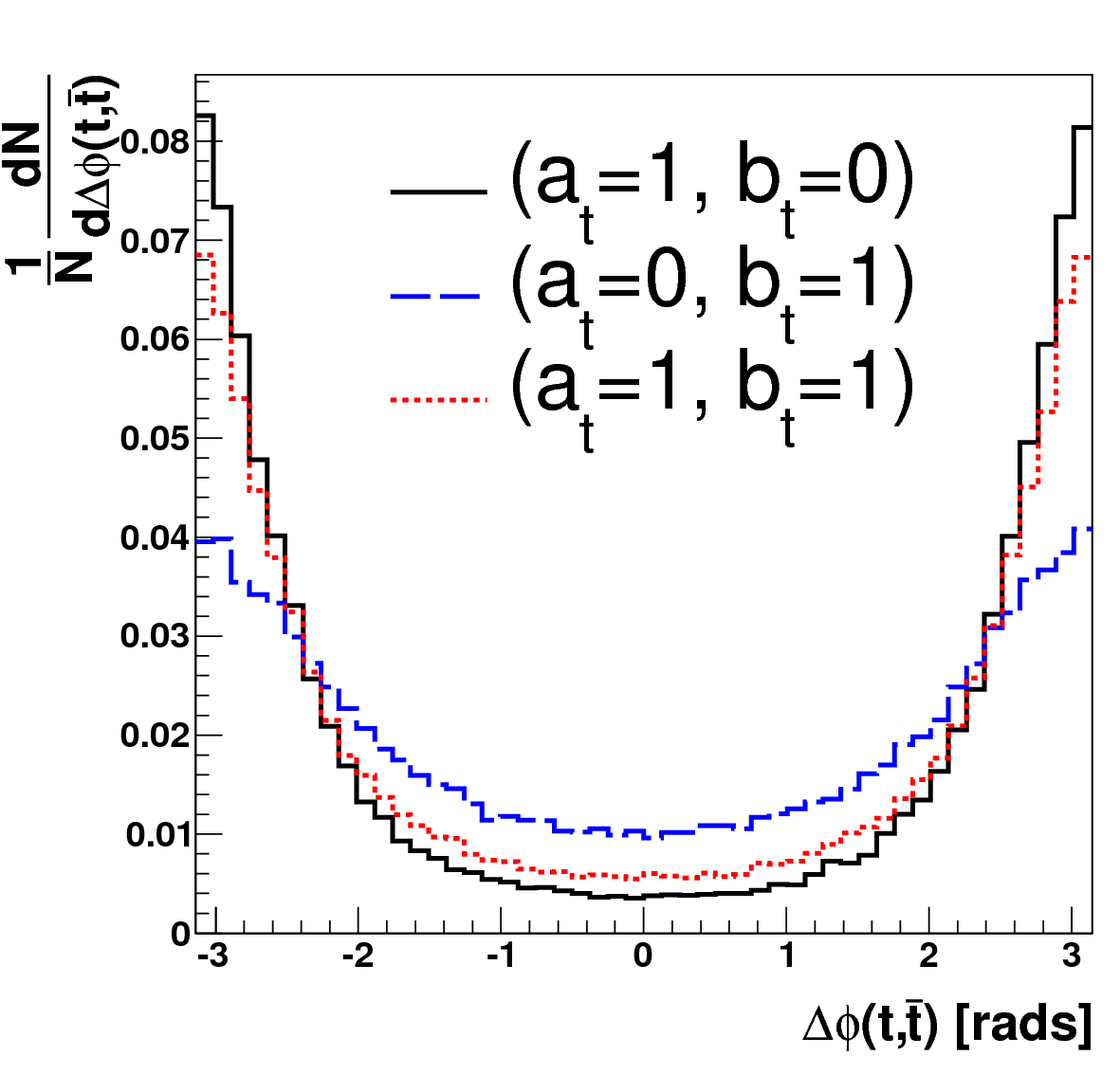}
\caption{
The distribution of the azimuthal-angle difference between the top pair ($\Delta\phi(t,\bar{t})$), 
normalized to unity. The SM  distribution ($\at=1,\bt=0$) is shown with a solid black line, the 
pseudo-scalar case ($\at=0,\bt=1$) with a blue dashed line, and the CP-violating case ($\at=1,\bt=1$) 
with a dotted red line.
\label{fig:dphitt}
}
\end{figure}
The larger transverse momentum of the Higgs in the pseudo-scalar case will have an effect on an 
observable that can be measured quite easily, namely the azimuthal-angle separation between the top quark 
and anti-quark, $\Delta\phi(t,\bar{t})$. In order to measure this quantity one needs only to reconstruct 
one of the top momenta at most. The distribution for this observable is shown in fig. \ref{fig:dphitt} 
for the SM ($\at=1,\bt=0$), the pseudo-scalar ($\at=0,\bt=1$), and the CP-violating case ($\at=1,\bt=1$). 
We see that in either case $\Delta\phi(t,\bar{t})$ peaks at large values $\pm \pi$. However, for the 
pseudo-scalar case the distribution is more flat in comparison to the SM. This can be understood as 
follows. For events produced near the energy threshold the transverse momentum of the Higgs is small. 
This means that the top pair will be produced mostly back to back. This accounts for the peaks observed 
at $|\Delta\phi(t,\bar{t})|=\pi$. Because the $\pt^{h}$ distribution in the pseudo-scalar case is pushed 
to larger values, this will give rise to a flatter distribution in $\Delta\phi(t,\bar{t})$.
Considering that the construction of this observable only requires information about the direction of 
the various decay products, it can be readily used in both the hadronic as well as semi-leptonic decay 
modes of the top quarks.
Uncertainties in the measurement of this observable are likely to be much reduced in comparison to 
$M_{\tth}$.

One may also attempt to address the question, which of the observables, $M_{\tth}$, $\pt^{h}$ or 
$\Delta\phi(t,\bar{t})$, better discriminates between scalar and pseudo-scalar production, although at 
the experimental level one may rather opt for reconstructing all three and use them in a multi-variate 
analysis. To answer this question we perform a likelihood analysis, akin to the one described in 
ref.~\cite{Godbole:2014cfa}. For the sake of comparison we assume $100\%$ efficiency in the construction 
of both observables, neglect backgrounds and normalize the total cross-section for scalar and 
pseudo-scalar production to be the same. As a result the luminosities that one will achieve from such an 
analysis are not realistic and are only to be used to appreciate the discriminating power of the two 
observables. We use histograms of the distributions binned with 20 intervals in the range $(0,2000)$~GeV, 
$(0,500)$~GeV and $(-\pi,\pi)$ for $M_{\tth}$, $\pt^{h}$  and $\Delta\phi(t,\bar{t})$, respectively.
\begin{figure}
\centering
\includegraphics[scale=0.10]{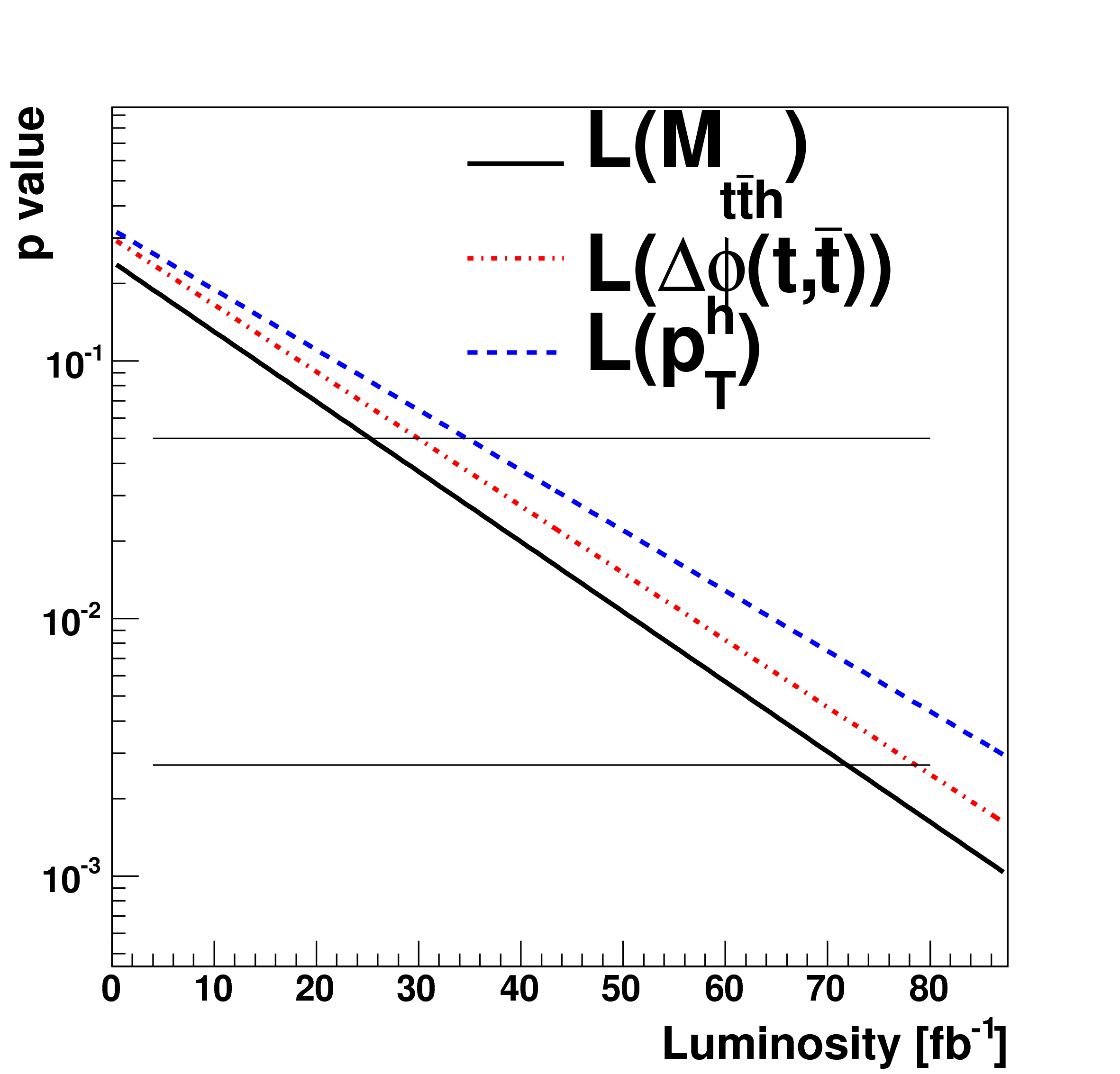}
\caption{
$p$-value as a function of the integrated luminosity for the distinction between the SM 
($\at=1,\bt=0$) and the pseudo-scalar case ($\at=0,\bt=1$) using likelihoods constructed from the 
observables $M_{\tth}$ (black solid line), $\pt^{h}$ (blue dashed line) and $\Delta\phi(t,\bar{t})$ 
(red dot-dashed line). We only include statistical uncertainties associated to the $\tth$ signal, in 
the absence of backgrounds. The two horizontal lines indicate the $2\sigma$ (top line) and $3\sigma$ 
(bottom line) exclusion limits.
\label{fig:mll}
}
\end{figure}

In fig. \ref{fig:mll}, we show the variation of the $p$-value for the pseudo-scalar hypothesis measured 
from the median value of the SM (null) hypothesis. We have used three likelihood functions, $L(M_{\tth})$,
$L(\pt^{h})$ and $L(\Delta\phi(t,\bar{t}))$. We reiterate that the absolute values of the luminosities in
this figure are not to be taken seriously, as we are only interested in the slopes of the lines.
From this figure we can infer that the $M_{\tth}$ distribution has a slightly better discriminating power
followed by $\Delta\phi(t,\bar{t})$ and then by $\pt^{h}$. 
However, the difference between the three likelihoods is very small. Since $\Delta\phi(t,\bar{t})$ and 
$\pt^{h}$ will have better reconstruction efficiencies and reduced uncertainties in comparison to 
$M_{\tth}$, the former are expected to perform much better in a more realistic analysis. We conclude that 
$\Delta\phi(t,\bar{t})$ and $\pt^{h}$ are better suited observables to distinguish between a scalar and a 
pseudo-scalar hypothesis.

So far we have only considered the kinematics of $\tth$ production, without any regards to the decays of 
the top quarks or the Higgs. Furthermore, the observables we have constructed are not directly sensitive 
to CP-violating effects. We will address these issues in the next section.

It is interesting to note that, although a specific measurement of the $t \bar t h$ cross section cannot 
discriminate between a scalar and a pseudo-scalar Higgs, the fact that the distributions are sensitive to 
its CP assignment means that by comparing a subset of the same cross section one could in principle lift 
the degeneracy. Normalised to the SM cross section the inclusive $t \bar t h$  cross section at 14 TeV can 
be written as 
\be
\sigma/\sigma^{\rm SM} \simeq a_t^2 + 0.42 \; b_t^2~.
\label{eq:incl14t}
\ee
A cut $\chi_{{\rm cut}}$, such as $p_T^h > 100$ GeV, increases the relative weight of the pseudo-scalar contribution
\be
\sigma(\chi_{{\rm cut}})/\sigma^{\rm{SM}}(\chi_{\rm cut})/=a_t^2 + 0.60\;b_t^2~.
\label{eq:incl14tcut}
\ee
If both these measurements, eq.~(\ref{eq:incl14t}) {\em and}  eq.~(\ref{eq:incl14tcut}), were precise 
enough, combining them could return non-zero values for {\em both} $a_t$ and $b_t$. While none of 
these cross sections is a measure of CP violation, the combination of both cross sections may lead non 
zero values for both $a_t$ and $b_t$, which is an indirect measure of CP violation. At the lower centre 
of mass energy of 8 TeV, the inclusive cross section benefits less the pseudo-scalar contribution. In 
fact, even at 14 TeV, the pseudo-scalar contribution is enhanced relative to the scalar contribution in 
the more energetic regions of phase space. The cross section at 8 TeV centre of mass is parametrised 
as 
\be
\sigma_{8{\rm TeV}}/\sigma^{\rm SM}_{8\,{\rm TeV}} \simeq a_t^2 + 0.31 \; b_t^2~.
\ee
It should be remembered at this point that precision of cross-section ratios as probes of BSM physics 
is to some extent limited by QCD uncertainties in the cross-section predictions~\cite{Baglio:2012et}. 

\subsection{Spin correlations in $\ttb$ decay products}

\begin{figure}
\centering
\includegraphics[scale=0.19]{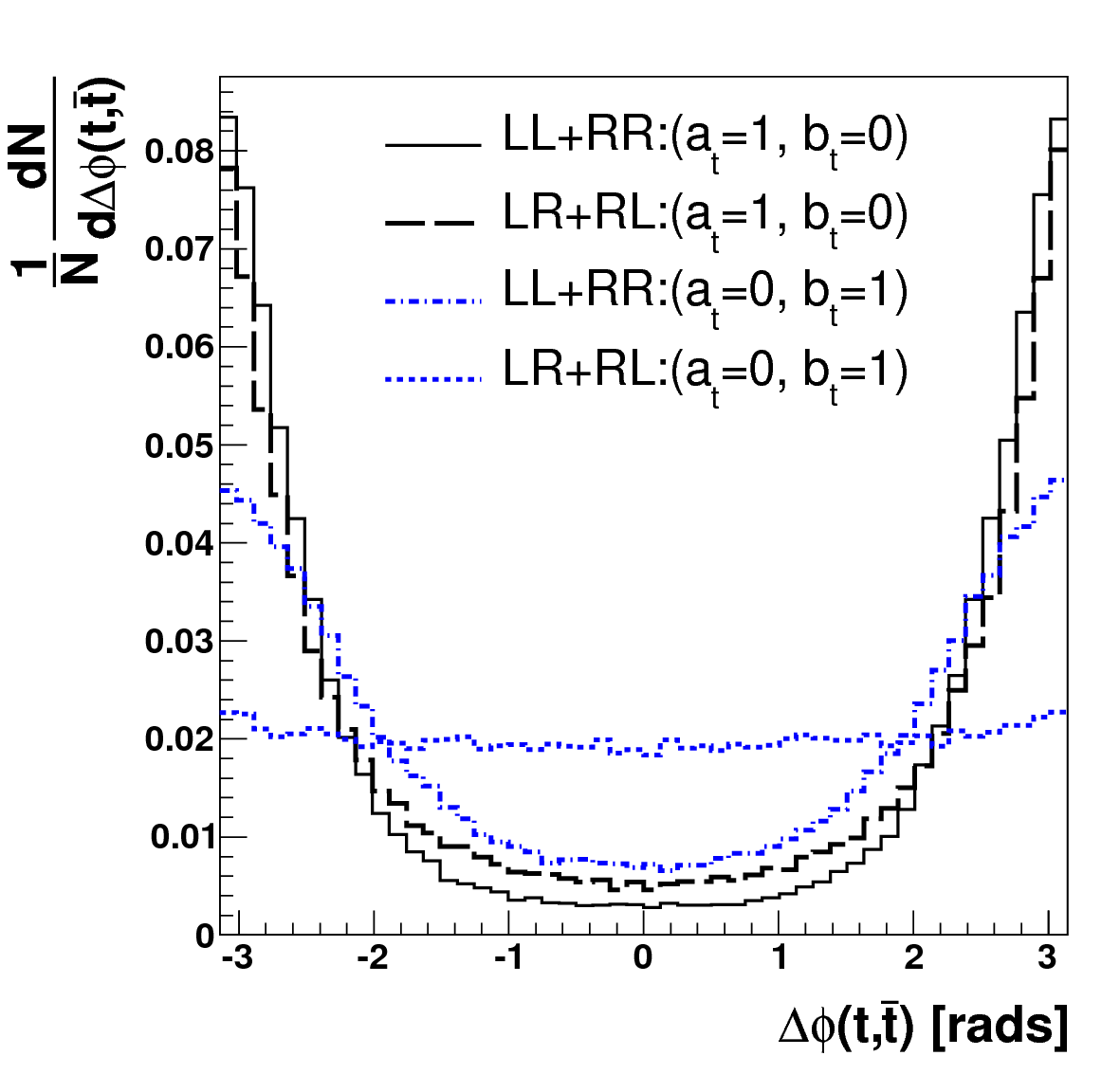}
\caption{
Distributions of the azimuthal-angle difference between the top pair, $\Delta\phi(t,\bar{t})$, 
normalized to unity. The four histograms refer to $\tth$ production with like-helicity top pairs 
($t_{L}\bar{t}_{L} +t_{R}\bar{t}_{R}$) and a scalar (black solid line) or a pseudo-scalar 
(blue dash-dotted line) Higgs, as well as to $\tth$ production with unlike-helicity top pairs 
($t_{L}\bar{t}_{R} +t_{R}\bar{t}_{L}$) and a scalar (black dashed line) or a pseudo-scalar 
(blue dotted line) Higgs, in the lab frame.
\label{fig:lrdphitt}
}
\end{figure}

The nature of the Higgs-top coupling in eq. (\ref{eq:ffh}) also affects spin correlations between 
the top and the anti-top quarks. The latter can be tested, for example, through azimuthal-angle 
differences between the momenta of the particles involved in the process \cite{Boudjema:2009fz,Buckley:2007th,Godbole:2006tq}.
We show in fig. \ref{fig:lrdphitt} the normalized distributions of $\Delta\phi(t,\bar{t})$ in 
unpolarized production for two helicity combinations of the final-state top quarks produced in 
association with a scalar or a pseudo-scalar Higgs. The two helicity combinations we consider are 
like-helicity ($t_{L}\bar{t}_{L} +t_{R}\bar{t}_{R}$) and unlike-helicity 
($t_{L}\bar{t}_{R} +t_{R}\bar{t}_{L}$) top pairs, in the lab frame. The conventions for helicity 
states and spinors are the same as in ref.~\cite{Murayama:1992gi}. 
The figure shows that the scalar and especially the pseudo-scalar cases produce different effects 
for different helicity combinations. The most striking difference occurs between the unlike-helicity 
combination for pseudo-scalar production, which yields a flat distribution, and the remaining 
distributions, all clearly peaked at $|\Delta\phi(t,\bar{t})|=\pi$.

A measure of the spin correlations can be defined through the following spin-correlation asymmetry
in the lab frame
\be
\label{eq:zeta_lab}
\zeta_{\rm lab} =
\frac{\sigma(pp \to t_{L}\bar{t}_{L} h) + \sigma(pp \to t_{R}\bar{t}_{R} h) 
- \sigma(pp \to t_{L}\bar{t}_{R} h) - \sigma(pp \to t_{R}\bar{t}_{L} h)}%
{\sigma(pp \to t_{L}\bar{t}_{L} h) + \sigma(pp \to t_{R}\bar{t}_{R} h) 
+ \sigma(pp \to t_{L}\bar{t}_{R} h) + \sigma(pp \to t_{R}\bar{t}_{L} h)}.
\ee
We find the following numerical values for the spin-correlation asymmetry for the different parity 
admixtures: $\zeta_{\rm lab}(\at=1,\bt=0)=0.22$, $\zeta_{\rm lab}(\at=0,\bt=1)=0.46$ and 
$\zeta_{\rm lab}(\at=1,\bt=1)=0.29$. These results can be combined in the following parametric formula
\be
\zeta _{\rm lab} \simeq \frac{0.22 \; \at^2 + 0.19 \; \bt^2}{\at^2+0.42 \; \bt^2}
\ee
valid for the case of the LHC at 14 TeV. The $a_t, b_t$ dependence of $\zeta_{\rm lab}$ confirms
our initial remark on the nature of the Higgs coupling affecting spin correlations. To be noted are 
the following points: 
{\em (i)} among the cases considered, the SM predicts the smallest value for $\zeta_{\rm lab}$; 
{\em (ii)} although the CP-violating case has a larger value for this coefficient, it is only marginally 
higher than the SM. This is due to the scalar cross-sections being larger than the pseudo-scalar ones;
{\em (iii)} the asymmetry in eq. (\ref{eq:zeta_lab}) is not sensitive to CP-violating effects as it is
a CP-even quantity. The same is also true for the observables described in the previous section. 
Note however that a measurement of $a_t$ and $b_t$ is nonetheless an indirect measure of CP violation. 
Theoretically a value for $\zeta$ which deviates from $0.22$ or $0.46$, corresponds to both $a_t$ and $b_t$ 
being non zero.

While the spin-correlation asymmetry in eq. (\ref{eq:zeta_lab}) may serve as a yardstick for the order 
of magnitude of the effects to be expected, it is not an easily measurable quantity at the LHC. 
Spin-correlation observables typically exploit the fact that the $\ttb$ spin information is passed on 
to the kinematic distributions of the decay products of the top quarks. In addition, the kinematics of 
the decay products are more likely to be affected by CP violation in the production process than the 
kinematics of the top quarks themselves, i.e. observables constructed using the decay products are more likely to be linearly sensitive to $\bt$.\footnote{In fact, CP-violating interference terms are more 
likely to be generated in the matrix element squared when we sum over helicities of the decay products 
since the matrix elements for production and decay can be linked through a density matrix.}

\begin{figure}
\centering
\includegraphics[scale=0.18]{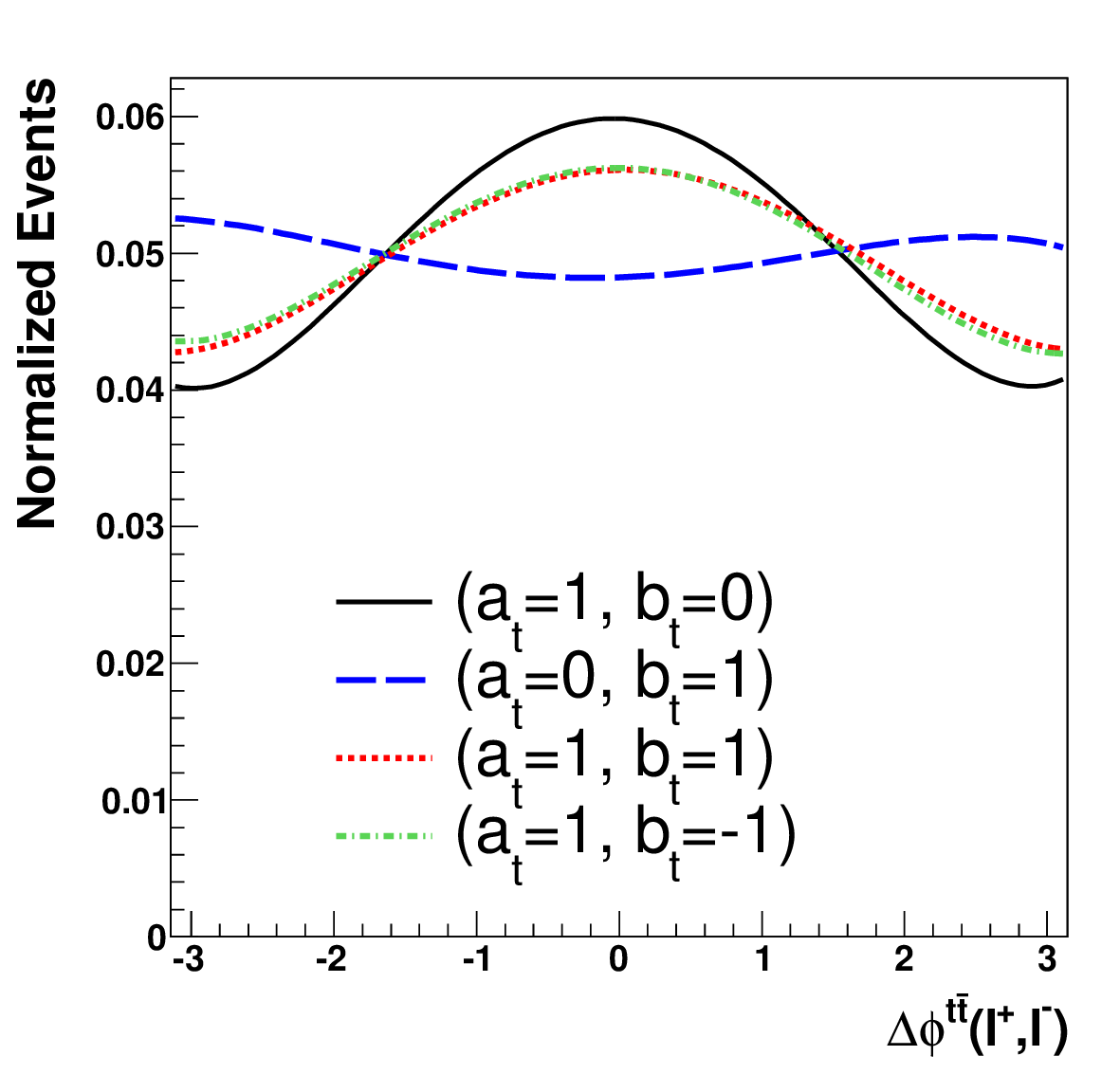}
\caption{
Normalized distributions for $\Delta\phi^{\ttb}(\ell^+,\ell^-)$ 
in $\tth$ production.
Distributions are shown for the SM $(\at=1,\bt=0)$ (black solid line), for the pseudo-scalar case
$(\at=0,\bt=1)$ (blue dashed line) and for two CP-violating cases, $(\at=1,\bt=1)$ (red dotted line) 
and $(\at=1,\bt=-1)$ (green dot-dashed line).
\label{fig:spc1}
}
\end{figure}

Let us first consider the di-leptonic decay mode\footnote{The observables that we will consider can be 
altered in an obvious way so that they can be used in the semi-leptonic or even hadronic decays.} of the
top pair.\footnote{For the di-leptonic channel we apply the following set of cuts: $\pt$ of jets 
$\textgreater\; 20$ GeV , $|\eta|$ of jets $\textless\; 5$ , $|\eta|$ of b jets $\textless \; 2.5$; 
$\pt$ of leptons $\textgreater\; 10$ GeV, $|\eta|$ of leptons $\textless\; 2.5$.} It is well known that 
the azimuthal-angle difference between the anti-lepton and the lepton from the decay of $t$ and $\bar t$, respectively, provides a good probe of spin-correlation effects in $\ttb$ production 
\cite{Godbole:2006tq,Mahlon:2010gw,Mahlon:1995zn,ATLAS:2012ao,CMS-PAS-TOP-12-004}, even in the lab frame. 
Furthermore, as the lepton angular distribution in the decay of the top is not affected by any 
non-standard effects in the decay vertex, it is a pure probe of
physics associated with the production process \cite{Godbole:2002qu,Godbole:2006tq}.
For $\tth$ production, the $t$ and $\bar{t}$ are not produced back to back (in the $xy$ plane) since the Higgs momentum adds an extra degree of freedom to the system. As a result spin-correlation effects in 
the azimuthal-angle difference will be washed out. 
It is possible to consider the angles between the two leptons in a different reference frame, where the 
kinematics of the $\tth$ system does not dissolve the effect of spin correlations. Distributions for 
such observables can be found in 
\cite{Heinemeyer:2013tqa,Biswas:2014hwa,Ellis:2013yxa,Demartin:2014fia,Brooijmans:2014eja}. 
In fig. \ref{fig:spc1} we show the distribution of one such angle, $\Delta\phi^{\ttb}(\ell^+,\ell^-)$~\cite{Brooijmans:2014eja,Heinemeyer:2013tqa,Demartin:2014fia}.
$\Delta\phi^{\ttb}(\ell^+,\ell^-)$ is defined as the difference between the azimuthal angle of the $\ell^+$ 
momentum in the rest frame of the top and the azimuthal angle of the $\ell^-$ momentum evaluated in the 
rest frame of the anti-top \cite{Heinemeyer:2013tqa,Brooijmans:2014eja}.\footnote{%
In constructing the $\ell^\pm$ momenta as described, we keep fixed for all events the choice of the $x$ 
and $y$ axes, and the $z$ axis is chosen, as customary, to lie along the beam direction. While individually
the azimuthal angles for the $\ell^+$ and $\ell^-$ momenta do depend on the choice of the $x$ and $y$
axes, their difference, as in $\Delta \phi$, does not. $\Delta \phi$ depends only on the choice of the 
beam axis. In fact, one can construct $\Delta \phi$ from the following formula
\be
\label{eq:cosdphi}
\cos(\Delta \phi^{\ttb}(\ell^+,\ell^-)) =
\frac{(\hat z \times {\vec p}^{~\bar t}_{\ell^-}) \cdot (\hat z \times {\vec p}^{~t}_{\ell^+})}
{|{\vec p}^{~\bar t}_{\ell^-}||{\vec p}^{~t}_{\ell^+}|}~,
\ee
that shows dependence only on the $\hat z$ direction. In this formula, the superscripts $t$ ($\bar t$) 
indicate that the given momentum is calculated in the rest frame of the $t$ ($\bar t$). 
}

From the figure we can see that the SM $(\at=1,\bt=0)$ 
distribution peaks at $\Delta\phi^{\ttb}(\ell^+,\ell^-)=0$, while the pseudo-scalar $(\at=0,\bt=1)$ case  
has a minimum at $\Delta\phi^{\ttb}(\ell^+,\ell^-)=0$. We have also considered two CP-violating 
cases, $(\at=1,\bt=1)$ and $(\at=1,\bt=-1)$, which show a behaviour qualitatively similar to the SM case. 
Furthermore, since the distributions for the two CP-violating cases appear to be the same, we can conclude 
that the two observables do not depend on $\bt$ linearly and hence do not probe CP violation in the 
production process in a direct manner.

\begin{figure}
\centering
\includegraphics[scale=0.2]{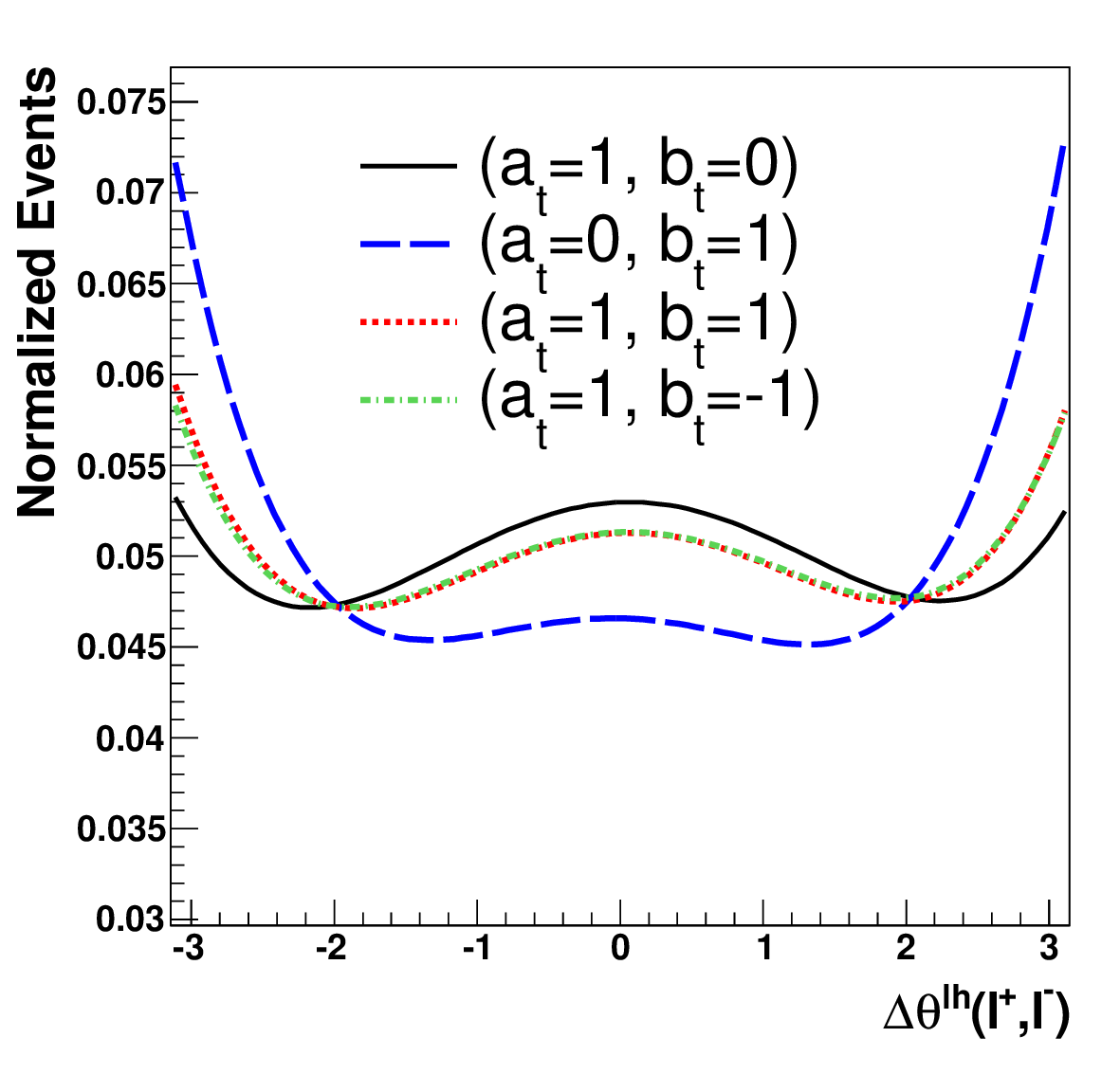}
\caption{
Normalized distributions for $\Delta\theta^{\ell h}(\ell^+,\ell^-)$ in $\tth$ production.
Distributions are shown for SM $(\at=1,\bt=0)$ (black solid line), for a pseudo-scalar $(\at=0,\bt=1)$ 
(blue dashed line)  and for two CP violating cases, $(\at=1,\bt=1)$ (red dotted line) and $(\at=1,\bt=-1)$ 
(green dot-dashed line).
\label{fig:dtheta^lh}
}
\end{figure}
Although the $\Delta\phi^{\ttb}(\ell^+,\ell^-)$ and other observables considered in the literature~\cite{Ellis:2013yxa} do manage to differentiate between a scalar and a pseudo-scalar, they are 
extremely difficult to construct at the LHC, especially because a full reconstruction of all momenta of 
the $\tth$ system is necessary. In addition, the uncertainties in the measurement of the various momenta
involved will carry over to the uncertainties in the measurement of these observables as we transform 
between different frames of reference. We therefore explore the option of constructing lab-frame 
observables. One such observable is $\Delta\theta^{\ell h}(\ell^+,\ell^-)$, defined as the angle between 
the two lepton momenta projected onto the plane perpendicular to the $h$ direction in the lab frame:
\be
\cos(\Delta\theta^{\ell h}(\ell^+,\ell^-)) = \frac{(\vec{p}_{h} \times {\vec p}_{\ell^+})
\cdot (\vec{p}_{h} \times {\vec p}_{\ell^-}) }
{|\vec{p}_{h} \times {\vec p}_{\ell^+}||\vec{p}_{h} \times {\vec p}_{\ell^-}|}~.
\label{eq:dtheta^lh}
\ee
This definition can be understood from the following argument. Recall that, for two-body $\ttb$ production, 
the azimuthal angle of the two leptons is sensitive to spin correlation effects. The $\tth$ system follows
three body kinematics, hence the $\ttb$ can be understood to `recoil' off the Higgs. It follows that, when
we project the two lepton momenta onto the plane perpendicular to the Higgs direction, the angle between 
them will also be sensitive to such spin-correlation effects.

The distribution for $\Delta\theta^{\ell h}(\ell^+,\ell^-)$ is shown in fig. \ref{fig:dtheta^lh}. 
From this plot we see that, similarly as for the angles considered before, there is an extremum at 
$\Delta\theta^{\ell h} (\ell^+,\ell^-)=0$ for all cases considered. The SM distribution displays
a pronounced peak at $\Delta\theta^{\ell h}(\ell^+,\ell^-)=0$, while the pseudo-scalar distribution 
is smaller and flatter in the whole region $[-\pi/2, +\pi/2]$, whereas it is larger at 
$|\Delta\theta^{\ell h}(\ell^+,\ell^-)|=\pi$. Hence this observable can be used to probe the CP nature 
of the $\tth$ interaction. On the other 
hand, being by its definition a CP-even observable, $\Delta\theta^{\ell h}(\ell^+,\ell^-)$ does not
distinguish between the two CP-violating cases $(\at=1,\bt=1)$ and $(\at=1,\bt=-1)$, that in fact have
exactly the same behaviour in fig. \ref{fig:dtheta^lh}. In this respect, it is worth noting explicitly 
that, while the plot in fig. \ref{fig:dtheta^lh} spans the range $[-\pi, \pi]$, 
$\Delta\theta^{\ell h}(\ell^+,\ell^-)$ is, according to eq. (\ref{eq:dtheta^lh}), defined only in the 
interval $[0, \pi]$. In order to assign a given event to the $[0, \pi]$ or to the $[-\pi, 0]$ interval, 
one needs an observable proportional to $\sin \Delta\theta^{\ell h}(\ell^+,\ell^-)$, for example 
${\rm sgn} (\vec p_h \cdot (\vec p_{\ell^+} \times \vec p_{\ell^-}))$, where `$\sgn$' indicates that we 
consider the sign of the term in brackets.

We conclude this section by noting that, albeit not explicitly shown here, other distributions
that, like the one in fig. \ref{fig:dtheta^lh}, are also able to distinguish the different vertex 
structures, would arise if we were to replace one or both of the lepton momenta in eq. 
(\ref{eq:dtheta^lh}) by $W$-boson momenta. Such distributions are useful in semi-leptonic or fully 
hadronic decays of the top pair.

\subsection{CP-violating observables}

So far we have confined ourselves to observables that are not sensitive to CP-violating effects. 
An observable sensitive to CP violation must be odd under CP transformations. Such quantities have 
been considered in the context of $e^{+}e^{-}$ colliders \cite{Atwood:2000tu,Godbole:2002qu,%
Godbole:2006tq,Godbole:2007uz}, and these results were exploited in the optimal-observable analysis of 
ref. \cite{Gunion:1996vv}. More recently, in the context of the LHC, a CP-odd observable was proposed 
in ref. \cite{Ellis:2013yxa} as follows
\be
\alpha \equiv \sgn \left(\vec{p}_{t}^{\ \ttb}\cdot(\vec{p}_{\ell^{-}}^{\ \ttb}
\times\vec{p}_{\ell^{+}}^{\ \ttb})\right).
\ee
Here the superscripts indicate that the corresponding momenta are constructed in the centre-of-mass
frame of the $\ttb$ system. Because of `$\sgn$', $\alpha$ can only take values of $\pm 1$. 

Although this observable is sensitive to CP violation linear in $\bt$, it suffers from the
same problem as before: it is very difficult to reconstruct at the LHC as all momenta of the $\tth$ 
system need to be determined. We suggest an alternative CP-odd observable that can be constructed 
{\em entirely} out of lab-frame quantities:
\be
\label{eq:beta}
\beta \equiv 
\sgn \left((\vec{p}_{b}-\vec{p}_{\bar{b}})\cdot(\vec{p}_{\ell^{-}}\times\vec{p}_{\ell^{+}})\right).
\ee
Note that, in order to correctly identify jets originating from a $b$ and $\bar{b}$ quark, one needs 
not reconstruct the top or anti-top momenta, of course. Various algorithms can be used to differentiate 
$b$- from $\bar{b}$-jets.\footnote{See for example refs.~\cite{Aad:2013uza,Nachman:2014qma,CMS:2012wcp,Krohn:2012fg} and references therein.}

\begin{figure}
\centering
\includegraphics[scale=0.2]{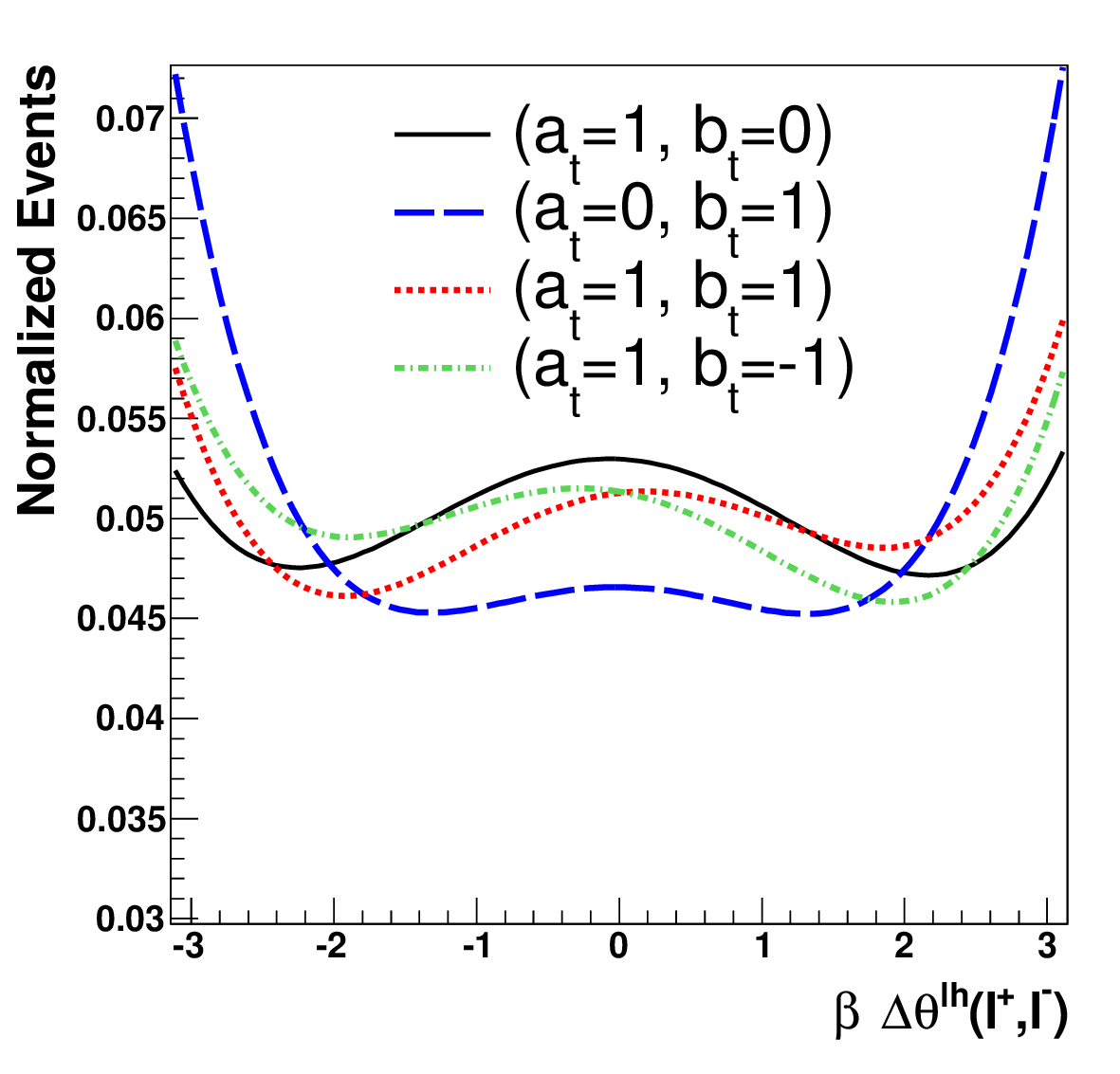}
\caption{
Normalized distributions for $\beta\cdot\Delta\theta^{\ell h}(\ell^+,\ell^-)$ in $\tth$ production.
Distributions are shown for the SM $(\at=1,\bt=0)$ (black solid line), for the pseudo-scalar case 
$(\at=0,\bt=1)$ (blue dashed line) and for two CP-violating cases, $(\at=1,\bt=1)$ (red dotted line) 
and $(\at=1,\bt=-1)$ (green dot-dashed line).
\label{fig:spc-cp2}
}
\end{figure}

The distribution obtained when we multiply $\beta$ by $\Delta\theta^{\ell h}(\ell^+,\ell^-)$ is shown in 
fig. \ref{fig:spc-cp2}. This distribution displays an asymmetry for the two CP-violating cases. 
Specifically, the distribution for $(\at=1,\bt=1)$ yields larger values in the positive $x$-axis, whereas 
the distribution for $(\at=1,\bt=-1)$ is larger on the negative $x$-axis.

We thus have a quantity that not only is sensitive to CP violation but is constructed entirely out of 
lab-frame kinematics. In addition, a measurement of this observable demands only reconstruction of the 
Higgs momentum, whereas reconstruction of the top pair momenta is not necessary.
Note on the other hand that this observable cannot be generalized easily to the case of semi-leptonic or 
hadronic decays of the top since it is not possible to differentiate between the quark and anti-quark jet 
originating from $W$-boson decays.

\begin{figure}
\centering
\includegraphics[scale=0.25]{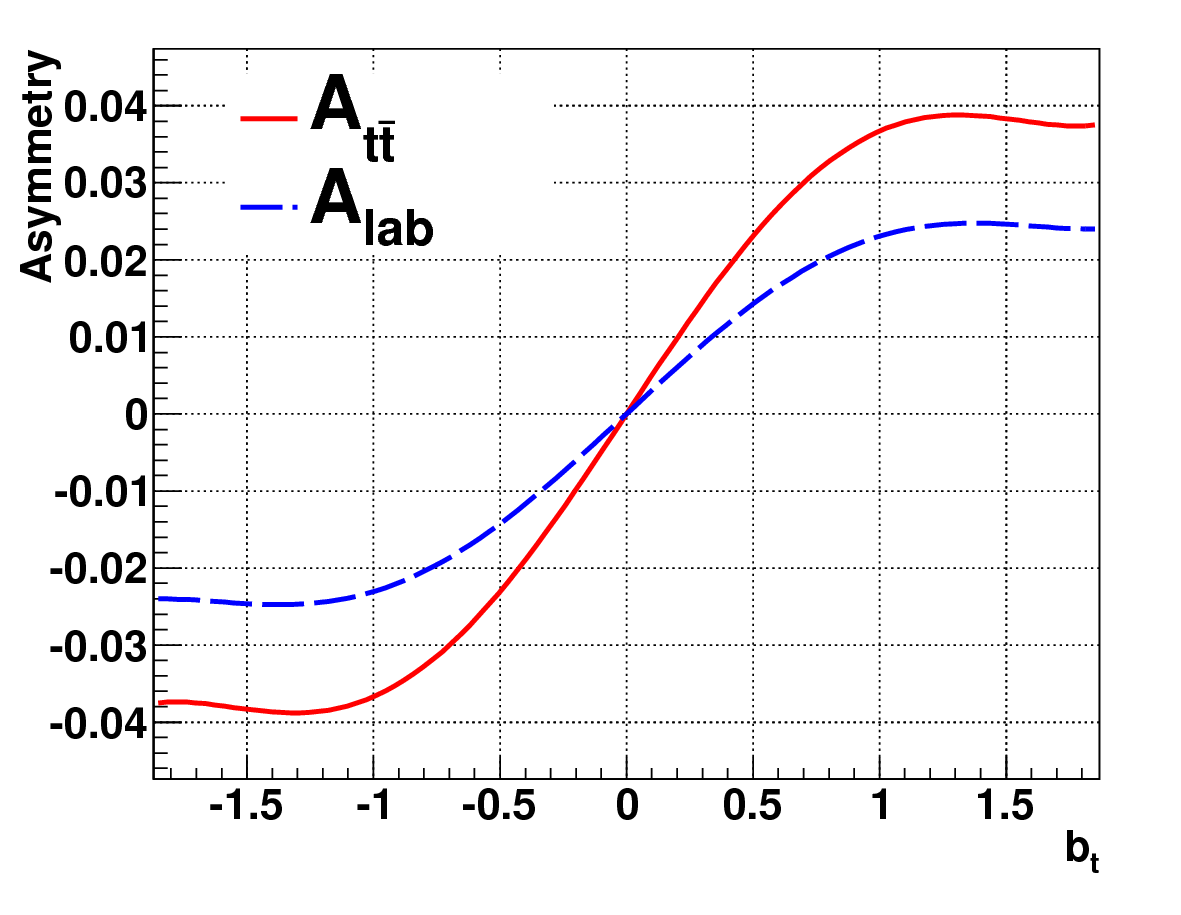}
\caption{
Dependence of the asymmetries $A_{\ttb}$ (red solid line) and $A_{\rm lab}$ (blue dashed line) on the 
strength of the CP-odd component $\bt$. In this plot $\at$ is kept fixed to unity.
\label{fig:asym}
}
\end{figure}

It is useful to define CP asymmetries with the observables 
$\alpha\times\Delta\theta^{\ttb}(\ell^+,\ell^-)$ \cite{Ellis:2013yxa} and 
$\beta\times\Delta\theta^{\ell h} (\ell^-,\ell^+)$ as follows
\be
A_{\ttb}=
\frac{\sigma(\alpha\times\Delta\theta^{\ttb}(\ell^+,\ell^-)>0)
-\sigma(\alpha\times\Delta\theta^{\ttb}(\ell^+,\ell^-)<0)}%
{\sigma(\alpha\times\Delta\theta^{\ttb}(\ell^+,\ell^-)>0)
+\sigma(\alpha\times\Delta\theta^{\ttb}(\ell^+,\ell^-)<0)}
\label{eq:as1}
\ee
and 
\be
A_{\rm lab}=\frac{\sigma(\beta\times\Delta\theta^{\ell h}(\ell^-,\ell^+) > 0)
-\sigma(\beta\times\Delta\theta^{\ell h}(\ell^-,\ell^+) < 0)}%
{\sigma(\beta\times\Delta\theta^{\ell h}(\ell^-,\ell^+) > 0)
+\sigma(\beta\times\Delta\theta^{\ell h}(\ell^-,\ell^+) < 0)}.
\label{eq:as2}
\ee
The dependence of these asymmetries on $\bt$ (keeping $\at=1$ fixed) is shown in fig. \ref{fig:asym}. 
We observe that both asymmetries are sensitive to the sign of $\bt$ (and hence linear in $\bt$), being 
negative for negative values of $\bt$ and positive for positive values of this parameter. The magnitude 
of the asymmetry $A_{\ttb}$ is larger than the magnitude of $A_{\rm lab}$ for a given value of $\bt$. 
However, we emphasize again that $A_{\rm lab}$ is constructed out of lab-frame quantities only and as 
such it is expected to be more easily measurable and to have less systematic uncertainties than 
$A_{\ttb}$.

\section{Conclusions} \label{sec:conclusions}

We have considered a Higgs-top Yukawa coupling that allows for a general scalar and pseudo-scalar Higgs
admixture, and have explored the possibility to probe this coupling in a model-independent framework. 
We find that, although constraining, the information provided by the Higgs rates, or by low-energy 
observables such as EDMs, does not suffice to provide conclusive evidence about the nature of this 
coupling. The arguably best way of probing this coupling unambiguously is its direct measurement. 
While certainly challenging in a hadronic environment like the LHC's, a measurement of this coupling 
would provide crucial information on the properties of the scalar coupled to the SM's heaviest particle, 
let alone the possibility of unveiling CP-violating effects. 

We have investigated some of the possible kinematic observables that could be used to discriminate a 
scalar- from a pseudo-scalar-like coupling at the LHC, focussing on the possibility of quantities 
constructed out of just {\em lab-frame} kinematics. The information about the nature of the coupling 
is carried by the threshold behaviour of the total invariant mass of the $\tth$ system, 
which is however very difficult to reconstruct. We find that similar information is encoded in the 
distributions of two experimentally simpler quantities, namely the transverse momentum of the Higgs 
and the azimuthal-angle separation between the $\ttb$ pair.

We furthermore exploit the fact that the information about the nature of the $\tth$ interaction is 
also passed on to the decay products of the $\ttb$ pair. Spin correlations between the $t$ and the 
$\bar t$ are likewise affected by the scalar vs. pseudo-scalar nature of this interaction.
We suggest several {\em lab-frame} observables that are affected by the spin correlations and hence 
can be used to probe the Higgs-top interactions in all possible decay modes of the $\ttb$ pair: 
di-leptonic, semi-leptonic and hadronic.

Finally, in the di-lepton channel we construct an observable that bears linear dependence on $\bt$ and 
hence is sensitive to CP-violating effects. We determine the corresponding CP asymmetry and show how 
it is sensitive to both the strength and the sign of $\bt$.

It goes without saying that, being an exploratory study aimed at the definition of lab-frame observables, 
the analysis performed here is simplistic. In particular it is a leading-order and parton-level analysis. 
While refinements towards a more realistic analysis (like inclusion of NLO, detector smearing, 
hadronization effects, etc.) will change quantitatively several of our distributions, they are 
not expected to modify our main conclusions. More detailed investigations are in progress.

\subsubsection*{Acknowledgments}

The authors acknowledge hospitality of the Centre de Physique des Houches, where this work started
in connection with the workshop PhysTeV 2013. We also thank Gustaaf Broijmaans for raising 
very useful questions. KM acknowledges the support of the Rh\^one-Alpes CMIRA programme and the excellence cluster Labex ENIGMASS as well as support by the National Science Foundation under Grant No. PHY-0854889. R.M.G. wishes to acknowledge support from the Department of Science and
Technology, India under Grant No. SR/S2/JCB-64/2007 under the J.C. Bose 
Fellowship scheme. R.M.G. also wishes to acknowledge support from the French ANR Project ``DMAstro-LHC'', 
 ANR-12-BS05-0006, for a visit to LAPTh.
\appendix

\section{Data used in fits} \label{sec:app1}

We present in the following tables \ref{c7:t1} to \ref{c7:t6} the data used in the fits.

\begin{table}[h!]
\centering
%\begin{ruledtabular}
\begin{tabular}{ c|c| c| c|c|c|c }
\hline
\hline
Channel & Signal strength $\mu$ & $m_h$(GeV) & 
\multicolumn{3}{c}{Production mode} \\ \hline
         &   &  & ggF & VBF & VH & ttH  \\
\hline
\multicolumn{7}{c}{ATLAS (4.5${\rm fb}^{-1}$ at 7TeV + 20.3 
fb$^{-1}$ at 8TeV)  \cite{Aad:2014eha} } \\
\hline
Inclusive  &$1.17\pm 0.23$ & 125.4 & 87.5\% & 7.1 \% & 4.9\% & 0.5\% 
    \\
\hline
\multicolumn{7}{c}{CMS (5.1${\rm fb}^{-1}$ at 7TeV + 19.7${\rm fb}^{-1}$ 
at 8TeV)  \cite{Khachatryan:2014ira} } \\
\hline
Inclusive  & $1.14^{+0.26}_{-0.23}$ & 124.7 
    & 87.5\% & 7.1\% & 4.9\% & 0.5\% \\
\hline
\multicolumn{7}{c}{Tevatron (10.0${\rm fb}^{-1}$ at 1.96TeV) \cite{Group:2012zca}}  \\
\hline
Combined & $6.14^{+3.25}_{-3.19}$ & 125 & 78\% & 5\% & 17\% & - \\
\hline
\end{tabular}
  \caption{ \label{c7:t1}
    Data on signal strengths of $h\rightarrow \gamma \gamma$
    recorded by ATLAS and CMS, and at the Tevatron 
    The percentages of each production mode in each data are given.
  }
%\end{ruledtabular}
\end{table}

\begin{table}[h!]
%\begin{ruledtabular}
\centering
\begin{tabular}{c |c| c| c|c|c|c }
\hline\hline
Channel & Signal strength $\mu$ & $m_h$(GeV) & 
\multicolumn{4}{c}{Production mode} 
\\ \hline
 &  & & ggF & VBF & VH & ttH  \\
\hline
\multicolumn{7}{c}
{ATLAS (4.8${\rm fb}^{-1}$ at 7TeV + 20.7${\rm fb}^{-1}$
  at 8TeV)  \cite{Aad:2014aba,Aad:2014eva}}\\
\hline
Inclusive &$1.66^{+0.45}_{-0.38} $ & 124.51 
   & 87.5\% & 7.1\% & 4.9\% & 0.5\%  \\
\hline
\multicolumn{7}{c}
{CMS (5.1${\rm fb}^{-1}$ at 7TeV + 19.6 ${\rm fb}^{-1}$ 
at 8TeV) \cite{Chatrchyan:2013mxa} }\\
\hline
Inclusive & $0.93^{+0.29}_{-0.25}$ & 125.6 
        & 87.5\% & 7.1\% & 4.9\% & 0.5\% \\
\hline
\end{tabular}
\caption{ \label{c7:t2}
Data on signal strengths of $\dhzz$ recorded by ATLAS and CMS. The percentages of each production mode in each data are given.
}
%\end{ruledtabular}
\end{table}

\begin{table}[h!]
\centering
%\begin{ruledtabular}
\begin{tabular}{c| c| c| c|c|c|c }
\hline \hline
Channel & Signal strength $\mu$ & $m_h$(GeV) & 
\multicolumn{4}{c}{Production mode} 
 \\ \hline
   & & & ggF & VBF & VH & ttH  \\
\hline
\multicolumn{7}{c}
{ATLAS (25${\rm fb}^{-1}$ integrated luminosity at 7 and 8 TeV.)  \cite{ATLAS-CONF-2014-060} }\\
\hline
ggF & $1.01^{+0.27}_{-0.25}$ & 125.36 
  & 100.0\% & - & - & -  \\
VBF & $1.28^{+0.53}_{-0.45}$ & 125.36 
   & - & 100\% & - & -  \\
\hline
\multicolumn{7}{c}
{CMS (up to 4.9 ${\rm fb}^{-1}$ at 7TeV + 19.4 
${\rm fb}^{-1}$ at 8TeV) \cite{Chatrchyan:2013iaa} }\\
\hline
0/1 jet  & $0.74^{+0.22}_{-0.20} $ & 125.6 & 97\% & 3\% & - & - 
 \\
VBF tag & $0.60^{+0.57}_{-0.46}$ & 125.6 & 17\% & 83\% & - & - \\
VH tag & $0.39^{+1.97}_{-1.87}$ & 125.6 & - & - & 100\% & - \\
WH tag & $0.56^{+1.27}_{-0.95}$ & 125.6 & - & - & 100\% & - \\
\hline
\multicolumn{7}{c}
{Tevatron (10.0${\rm fb}^{-1}$ at 1.96TeV)  \cite{Group:2012zca} }\\
\hline
Combined   &$0.85^{+0.88}_{-0.81}$ & 125 & 78\% & 5\% & 17\% & - \\
\hline
\end{tabular}
\caption{\label{c7:t3}
    Data on signal strengths of $\dhww$
    recorded by ATLAS, CMS and Tevatron. 
    The percentages of each production mode in each data are given.}
%\end{ruledtabular}
\end{table}
\begin{table}[h!]
\centering
%\begin{ruledtabular}
\begin{tabular}{c|c|c|c|c|c|c}
\hline\hline
Channel & Signal strength $\mu$ & $m_h$(GeV) & 
 \multicolumn{4}{c}{Production mode}  \\ \hline
        &      &            & ggF & VBF & VH & ttH \\
\hline
\multicolumn{6}{c}
{ATLAS (4.7${\rm fb}^{-1}$ at 7TeV + 20.3${\rm fb}^{-1}$ at 8TeV)
   \cite{Aad:2014xzb}}\\
\hline
VH tag & $0.52\pm 0.4$ & 125.36 & - & - & 100\% & -  \\
\hline
\multicolumn{6}{c}
{CMS (up to 5.1${\rm fb}^{-1}$ at 7TeV + 18.9${\rm fb}^{-1}$ at 8TeV)
 \cite{Chatrchyan:2013zna} }\\
\hline
VH tag & $1.0\pm 0.5$ & 125.8 & - & - & 100\% & -  \\
%ttH tag & $-0.80^{+2.10}_{-1.84}$ & 125.8 & - & - & - & 100\% \\
\hline
\multicolumn{6}{c}
{Tevatron (10.0${\rm fb}^{-1}$ at 1.96TeV) \cite{Group:2012zca} }\\
\hline
VH tag & $1.56^{+0.72}_{-0.73}$ & 125 & - & - & 100\% & -  \\
\hline
\end{tabular}
%\end{ruledtabular}
\caption{\label{c7:t4}
    Data on signal strengths of $h\to b \bar{b}$
    recorded by ATLAS, CMS and Tevatron. 
    The percentages of each production mode in each data are given.}
\end{table}
\begin{table}[h!]
\centering
\begin{tabular}{c| c| c| c|c|c|c }
\hline\hline
Channel & Signal strength $\mu$ & $m_h$(GeV) & 
\multicolumn{4}{c}{Production mode} 
 \\ \hline
  &  && ggF & VBF & VH & ttH  \\
\hline
\multicolumn{7}{c}
{ATLAS (4.5${\rm fb}^{-1}$ at 7TeV + 20.3${\rm fb}^{-1}$ at 8TeV)
\cite{ATLAS-CONF-2014-061}}\\
\hline
$\mu(ggF)$ & $1.93^{+1.42}_{-1.11}$ & 125.36 
 & 100\% & - & - & - \\
$\mu(VBF+VH)$  & $1.24^{+0.57}_{-0.53}$ & 125.36 & - 
  & 59.4\% & 40.6\%   \\
\hline
\multicolumn{7}{c}
{CMS (up to 4.9${\rm fb}^{-1}$ at 7TeV + 19.7 
${\rm fb}^{-1}$ at 8TeV) \cite{Chatrchyan:2014nva} }\\
\hline
0 jet  & $0.34\pm 1.09$ & 125 & 
96.9\% & 1.0\% & 2.1\% & -  \\
1 jet  & $1.07\pm 0.46$ & 125 & 
75.7\% & 14.0\% & 10.3\% & -  \\
VBF tag  & $0.95 \pm 0.41 $ & 125 & 
19.6\%  & 80.4\% & - & -  \\
VH tag  & $-0.33 \pm 1.02$ & 125 & - & - & 100\% &
  -  \\
\hline
\end{tabular}
\caption{
\label{c7:t5}
Data on signal strengths of $h\to \tau^{+} \tau^{-}$
recorded by ATLAS and CMS. 
The percentages of each production mode in each data are given.
For ATLAS data we use a correlation of $\rho=-0.5$.
 }
\end{table}

\begin{table}[h!]
\centering
\begin{tabular}{c| c| c| c|c|c|c }
\hline\hline
Channel & Signal strength $\mu$ & $m_h$(GeV) & 
\multicolumn{4}{c}{Production mode} 
 \\ \hline
  &  && ggF & VBF & VH & ttH  \\
\hline
\multicolumn{7}{c}
{ATLAS (4.5${\rm fb}^{-1}$ at 7TeV + 20.3${\rm fb}^{-1}$ at 8TeV)
\cite{Aad:2014lma}}\\
\hline
$\gamma \gamma$  & $1.3^{2.62}_{-1.72}$ & 125.4 & - & - & - & 100\% \\
\hline
\multicolumn{7}{c}
{ATLAS ( 20.3${\rm fb}^{-1}$ at 8TeV)
\cite{Aad:2015gra}}\\
\hline
$b\bar{b}$  & $+1.5^{+1.1}_{-1.1}$ & 125 & - & - & - & 100\% \\
\hline
\multicolumn{7}{c}
{CMS (up to 5.1${\rm fb}^{-1}$ at 7TeV + 19.7 
${\rm fb}^{-1}$ at 8TeV) \cite{Khachatryan:2014qaa} }\\
\hline
$\gamma \gamma$  & $2.7^{2.6}_{-1.8}$ & 125.6 & - & - & - & 100\% \\
$b\bar{b}$  & $+0.7^{+1.9}_{-1.9}$ & 125.6 & - & - & - & 100\% \\
$\tau_{h}\tau_{h}$  & $-1.3^{+6.3}_{-5.5} $ & 125.6 & - & - & - & 100\% \\
4-lepton  & $-4.7^{+5.0}_{-1.3} $ & 125.6 & - & - & - & 100\% \\
3-lepton  & $+3.1^{+2.4}_{-2.0} $ & 125.6 & - & - & - & 100\% \\
Same-sign 2l  & $+5.3^{+2.1}_{-1.8} $ & 125.6 & - & - & - & 100\% \\
\hline
\multicolumn{6}{c}
{ATLAS ( 20.3  
	${\rm fb}^{-1}$ at 8TeV) \cite{ATLAS-CONF-2015-006} }\\
\hline
$2\text{lepton}0\tau_{had}$  & $2.8^{2.1}_{-1.9}$ & 125.0 & - & - & - & 100\% \\
3-lepton  & $+2.8^{+2.2}_{-1.8}$ & 125.0 & - & - & - & 100\% \\
$2\text{lepton}1\tau_{had}$  & $-0.9^{+3.1}_{-2.0} $ & 125.0 & - & - & - & 100\% \\
4-lepton  & $1.8^{+6.9}_{-2.0} $ & 125.0 & - & - & - & 100\% \\
$1\text{lepton}2\tau_{had}$  & $-9.6^{+9.6}_{-9.7} $ & 125.0 & - & - & - & 100\% \\
\hline
\end{tabular}
\caption{
\label{c7:t6}
Data on signal strengths for various decay modes of the Higgs which is produced through the $\tth$ production mode, for both ATLAS and CMS. Note that, in the various analyses, contaminations from other production modes are negligible.
 }
\end{table}

\clearpage

\section{Form factors} \label{sec:app2}
The loop functions that we have used in the text appear in the analytic expressions for the one-loop 
induced widths $h \to \gamma \gamma$ and $h \to gg$. Keeping only the dominant $W$ and top contributions 
we have (see \cite{Djouadi:2005gj,Spira:1995rr})
\begin{eqnarray}
\label{eq:dec_app}
\Gamma_{\gamma \gamma} &=& 
\frac{G_F \alpha^2 m_H^3}{128 \sqrt{2} \pi^3} 
\bigg\{ \left|  
\kappa_W \, A_W^{a} (\tau_W) + \frac{4}{3} a_t \,  
A_t^{a} (\tau_t) \right|^2 + \left| \frac{4}{3} \bt \, A^b_t(\tau_t) \right|^2 \bigg\}~, \nn \\
\Gamma_{g g} &=& 
\frac{G_F \alpha_s^2 m_H^3}{64 \sqrt{2} \pi^3}
\bigg\{ \Bigl|  a_t \, \, 
A_t^{a} (\tau_t) \Bigl|^2 + 
\Bigl|  \bt \,  \, 
A_t^{b} (\tau_t) \Bigl|^2 \bigg\}~,
\end{eqnarray}
where
\bea
\label{eq:A_amplitudes}
A^a_t(\tau)&=&\frac{2}{\tau^2}(\tau + (\tau -1)f(\tau)), \\ \nn
A^a_W(\tau)&=&-\frac{1}{\tau^2}(2\tau^2 +3\tau +3(2\tau -1)f(\tau)), \\ \nn
A^b_t(\tau)&=&\frac{2}{\tau}f(\tau),
\eea
with $\tau_i=\frac{m_h^2}{4m_i^2}$ and
\bea
f (\tau) = \left\{
\begin{array}{lc}
\mbox{arcsin}^2 \sqrt{\tau}  & \mbox{for }\tau \leq 1 \\
- \frac{1}{4} \left[ \log \frac{1+\sqrt{1-\tau^{-1}}}{1-\sqrt{1-\tau^{-1}}} - i \pi \right]^2 & 
\mbox{for }\tau > 1
\end{array} \right.\,.
\eea

Since $\tau_t \ll 1$, an expansion of the loop functions $A_t^{a,b}$ in $\tau_t$ confirms that the 
departure from the infinite mass limit is very small. Indeed we can write
\bea
A_t^a(\tau_t) &=& \frac{4}{3} \left( 1+\frac{7}{30}\tau_t + {\cal {O}}(\tau_t^2) \right), \nn \\
A_t^b(\tau_t) &=& 2 \left( 1+\frac{\tau_t}{3} + {\cal {O}}(\tau_t^2) \right)~.
\eea
With $\tau_t = m_h^2 / (4 m_t^2) \simeq 0.15$, one sees that corrections to the amplitudes in the 
$\tau_t \to 0$ limit are respectively of order 3.5\% and 5\%.

\bibliographystyle{JHEP}
\bibliography{tth}

\providecommand{\href}[2]{#2}\begingroup\raggedright\begin{thebibliography}{100}

\bibitem{Aad:2012tfa}
{\bf ATLAS} Collaboration, G.~Aad {\em et.~al.}, {\it {Observation of a new
  particle in the search for the Standard Model Higgs boson with the ATLAS
  detector at the LHC}},  {\em Phys.Lett.} {\bf B716} (2012) 1--29,
  [\href{http://xxx.lanl.gov/abs/1207.7214}{{\tt arXiv:1207.7214}}].

\bibitem{Chatrchyan:2012ufa}
{\bf CMS} Collaboration, S.~Chatrchyan {\em et.~al.}, {\it {Observation of a
  new boson at a mass of 125 GeV with the CMS experiment at the LHC}},  {\em
  Phys.Lett.} {\bf B716} (2012) 30--61,
  [\href{http://xxx.lanl.gov/abs/1207.7235}{{\tt arXiv:1207.7235}}].

\bibitem{Chatrchyan:2013lba}
{\bf CMS} Collaboration, S.~Chatrchyan {\em et.~al.}, {\it {Observation of a
  new boson with mass near 125 GeV in pp collisions at $\sqrt{s}$ = 7 and 8
  TeV}},  {\em JHEP} {\bf 1306} (2013) 081,
  [\href{http://xxx.lanl.gov/abs/1303.4571}{{\tt arXiv:1303.4571}}].

\bibitem{Aad:2013xqa}
{\bf ATLAS} Collaboration, G.~Aad {\em et.~al.}, {\it {Evidence for the spin-0
  nature of the Higgs boson using ATLAS data}},  {\em Phys.Lett.} {\bf B726}
  (2013) 120--144, [\href{http://xxx.lanl.gov/abs/1307.1432}{{\tt
  arXiv:1307.1432}}].

\bibitem{ATLAS-CONF-2013-013}
{\bf ATLAS Collaboration} Collaboration, {\it {Measurements of the properties
  of the Higgs-like boson in the four lepton decay channel with the ATLAS
  detector using 25 fb?1 of proton-proton collision data}},  Tech. Rep.
  ATLAS-CONF-2013-013, CERN, Geneva, Mar, 2013.

\bibitem{Chatrchyan:2013mxa}
{\bf CMS Collaboration} Collaboration, S.~Chatrchyan {\em et.~al.}, {\it
  {Measurement of the properties of a Higgs boson in the four-lepton final
  state}},  {\em Phys.Rev.} {\bf D89} (2014) 092007,
  [\href{http://xxx.lanl.gov/abs/1312.5353}{{\tt arXiv:1312.5353}}].

\bibitem{Chatrchyan:2012jja}
{\bf CMS Collaboration} Collaboration, S.~Chatrchyan {\em et.~al.}, {\it {Study
  of the Mass and Spin-Parity of the Higgs Boson Candidate Via Its Decays to Z
  Boson Pairs}},  {\em Phys.Rev.Lett.} {\bf 110} (2013) 081803,
  [\href{http://xxx.lanl.gov/abs/1212.6639}{{\tt arXiv:1212.6639}}].

\bibitem{Choi:2002jk}
S.~Choi, D.~Miller, M.~Muhlleitner, and P.~Zerwas, {\it {Identifying the Higgs
  spin and parity in decays to Z pairs}},  {\em Phys.Lett.} {\bf B553} (2003)
  61--71, [\href{http://xxx.lanl.gov/abs/hep-ph/0210077}{{\tt
  hep-ph/0210077}}].

\bibitem{Godbole:2007cn}
R.~M. Godbole, D.~Miller, and M.~M. Muhlleitner, {\it {Aspects of CP violation
  in the H ZZ coupling at the LHC}},  {\em JHEP} {\bf 0712} (2007) 031,
  [\href{http://xxx.lanl.gov/abs/0708.0458}{{\tt arXiv:0708.0458}}].

\bibitem{Gao:2010qx}
Y.~Gao, A.~V. Gritsan, Z.~Guo, K.~Melnikov, M.~Schulze, {\em et.~al.}, {\it
  {Spin determination of single-produced resonances at hadron colliders}},
  {\em Phys.Rev.} {\bf D81} (2010) 075022,
  [\href{http://xxx.lanl.gov/abs/1001.3396}{{\tt arXiv:1001.3396}}].

\bibitem{DeRujula:2010ys}
A.~De~Rujula, J.~Lykken, M.~Pierini, C.~Rogan, and M.~Spiropulu, {\it {Higgs
  look-alikes at the LHC}},  {\em Phys.Rev.} {\bf D82} (2010) 013003,
  [\href{http://xxx.lanl.gov/abs/1001.5300}{{\tt arXiv:1001.5300}}].

\bibitem{Stolarski:2012ps}
D.~Stolarski and R.~Vega-Morales, {\it {Directly Measuring the Tensor Structure
  of the Scalar Coupling to Gauge Bosons}},  {\em Phys.Rev.} {\bf D86} (2012)
  117504, [\href{http://xxx.lanl.gov/abs/1208.4840}{{\tt arXiv:1208.4840}}].

\bibitem{Bolognesi:2012mm}
S.~Bolognesi, Y.~Gao, A.~V. Gritsan, K.~Melnikov, M.~Schulze, {\em et.~al.},
  {\it {On the spin and parity of a single-produced resonance at the LHC}},
  {\em Phys.Rev.} {\bf D86} (2012) 095031,
  [\href{http://xxx.lanl.gov/abs/1208.4018}{{\tt arXiv:1208.4018}}].

\bibitem{Chen:2014ona}
Y.~Chen, A.~Falkowski, I.~Low, and R.~Vega-Morales, {\it {New Observables for
  CP Violation in Higgs Decays}},
  \href{http://xxx.lanl.gov/abs/1405.6723}{{\tt arXiv:1405.6723}}.

\bibitem{Chatrchyan:2013iaa}
{\bf CMS Collaboration} Collaboration, S.~Chatrchyan {\em et.~al.}, {\it
  {Measurement of Higgs boson production and properties in the WW decay channel
  with leptonic final states}},  {\em JHEP} {\bf 1401} (2014) 096,
  [\href{http://xxx.lanl.gov/abs/1312.1129}{{\tt arXiv:1312.1129}}].

\bibitem{Khachatryan:1969386}
{\bf CMS Collaboration} Collaboration, V.~e.~a. Khachatryan, {\it {Constraints
  on the spin-parity and anomalous $\mathrm{HVV}$ couplings of the Higgs boson
  in proton collisions at 7 and 8 TeV}},  Tech. Rep. CERN-PH-EP-2014-265.
  CMS-HIG-14-018-003. arXiv:1411.3441, CERN, Geneva, Nov, 2014.
\newblock Comments: Submitted to Phys. Rev. D.

\bibitem{Dawson:2013bba}
S.~Dawson, A.~Gritsan, H.~Logan, J.~Qian, C.~Tully, {\em et.~al.}, {\it
  {Working Group Report: Higgs Boson}},
  \href{http://xxx.lanl.gov/abs/1310.8361}{{\tt arXiv:1310.8361}}.

\bibitem{Plehn:2001nj}
T.~Plehn, D.~L. Rainwater, and D.~Zeppenfeld, {\it {Determining the structure
  of Higgs couplings at the LHC}},  {\em Phys.Rev.Lett.} {\bf 88} (2002)
  051801, [\href{http://xxx.lanl.gov/abs/hep-ph/0105325}{{\tt
  hep-ph/0105325}}].

\bibitem{Hankele:2006ma}
V.~Hankele, G.~Klamke, D.~Zeppenfeld, and T.~Figy, {\it {Anomalous Higgs boson
  couplings in vector boson fusion at the CERN LHC}},  {\em Phys.Rev.} {\bf
  D74} (2006) 095001, [\href{http://xxx.lanl.gov/abs/hep-ph/0609075}{{\tt
  hep-ph/0609075}}].

\bibitem{Andersen:2010zx}
J.~R. Andersen, K.~Arnold, and D.~Zeppenfeld, {\it {Azimuthal Angle
  Correlations for Higgs Boson plus Multi-Jet Events}},  {\em JHEP} {\bf 1006}
  (2010) 091, [\href{http://xxx.lanl.gov/abs/1001.3822}{{\tt
  arXiv:1001.3822}}].

\bibitem{Andersen:2008gc}
J.~R. Andersen, V.~Del~Duca, and C.~D. White, {\it {Higgs Boson Production in
  Association with Multiple Hard Jets}},  {\em JHEP} {\bf 0902} (2009) 015,
  [\href{http://xxx.lanl.gov/abs/0808.3696}{{\tt arXiv:0808.3696}}].

\bibitem{Djouadi:2013yb}
A.~Djouadi, R.~Godbole, B.~Mellado, and K.~Mohan, {\it {Probing the spin-parity
  of the Higgs boson via jet kinematics in vector boson fusion}},  {\em
  Phys.Lett.} {\bf B723} (2013) 307--313,
  [\href{http://xxx.lanl.gov/abs/1301.4965}{{\tt arXiv:1301.4965}}].

\bibitem{Englert:2012xt}
C.~Englert, D.~Goncalves-Netto, K.~Mawatari, and T.~Plehn, {\it {Higgs Quantum
  Numbers in Weak Boson Fusion}},  {\em JHEP} {\bf 1301} (2013) 148,
  [\href{http://xxx.lanl.gov/abs/1212.0843}{{\tt arXiv:1212.0843}}].

\bibitem{Han:2009ra}
T.~Han and Y.~Li, {\it {Genuine CP-odd Observables at the LHC}},  {\em
  Phys.Lett.} {\bf B683} (2010) 278--281,
  [\href{http://xxx.lanl.gov/abs/0911.2933}{{\tt arXiv:0911.2933}}].

\bibitem{Christensen:2010pf}
N.~D. Christensen, T.~Han, and Y.~Li, {\it {Testing CP Violation in ZZH
  Interactions at the LHC}},  {\em Phys.Lett.} {\bf B693} (2010) 28--35,
  [\href{http://xxx.lanl.gov/abs/1005.5393}{{\tt arXiv:1005.5393}}].

\bibitem{Desai:2011yj}
N.~Desai, D.~K. Ghosh, and B.~Mukhopadhyaya, {\it {CP-violating HWW couplings
  at the Large Hadron Collider}},  {\em Phys.Rev.} {\bf D83} (2011) 113004,
  [\href{http://xxx.lanl.gov/abs/1104.3327}{{\tt arXiv:1104.3327}}].

\bibitem{Ellis:2012xd}
J.~Ellis, D.~S. Hwang, V.~Sanz, and T.~You, {\it {A Fast Track towards the
  `Higgs' Spin and Parity}},  {\em JHEP} {\bf 1211} (2012) 134,
  [\href{http://xxx.lanl.gov/abs/1208.6002}{{\tt arXiv:1208.6002}}].

\bibitem{Godbole:2014cfa}
R.~M. Godbole, D.~J. Miller, K.~A. Mohan, and C.~D. White, {\it {Jet
  substructure and probes of CP violation in Vh production}},
  \href{http://xxx.lanl.gov/abs/1409.5449}{{\tt arXiv:1409.5449}}.

\bibitem{Godbole:2013saa}
R.~Godbole, D.~J. Miller, K.~Mohan, and C.~D. White, {\it {Boosting Higgs CP
  properties via VH Production at the Large Hadron Collider}},  {\em
  Phys.Lett.} {\bf B730} (2014) 275--279,
  [\href{http://xxx.lanl.gov/abs/1306.2573}{{\tt arXiv:1306.2573}}].

\bibitem{Delaunay:2013npa}
C.~Delaunay, G.~Perez, H.~de~Sandes, and W.~Skiba, {\it {Higgs Up-Down CP
  Asymmetry at the LHC}},  {\em Phys.Rev.} {\bf D89} (2014) 035004,
  [\href{http://xxx.lanl.gov/abs/1308.4930}{{\tt arXiv:1308.4930}}].

\bibitem{Berge:2014sra}
S.~Berge, W.~Bernreuther, and S.~Kirchner, {\it {Determination of the Higgs CP
  mixing angle in the tau decay channels at the LHC including the Drell-Yan
  background}},  {\em Eur.Phys.J.} {\bf C74} (2014), no.~11 3164,
  [\href{http://xxx.lanl.gov/abs/1408.0798}{{\tt arXiv:1408.0798}}].

\bibitem{Berge:2012wm}
S.~Berge, W.~Bernreuther, and H.~Spiesberger, {\it {Determination of the CP
  parity of Higgs bosons in their tau decay channels at the ILC}},
  \href{http://xxx.lanl.gov/abs/1208.1507}{{\tt arXiv:1208.1507}}.

\bibitem{Berge:2011ij}
S.~Berge, W.~Bernreuther, B.~Niepelt, and H.~Spiesberger, {\it {How to pin down
  the CP quantum numbers of a Higgs boson in its tau decays at the LHC}},  {\em
  Phys.Rev.} {\bf D84} (2011) 116003,
  [\href{http://xxx.lanl.gov/abs/1108.0670}{{\tt arXiv:1108.0670}}].

\bibitem{Berge:2008wi}
S.~Berge, W.~Bernreuther, and J.~Ziethe, {\it {Determining the CP parity of
  Higgs bosons at the LHC in their tau decay channels}},  {\em Phys.Rev.Lett.}
  {\bf 100} (2008) 171605, [\href{http://xxx.lanl.gov/abs/0801.2297}{{\tt
  arXiv:0801.2297}}].

\bibitem{Djouadi:2013qya}
A.~Djouadi and G.~Moreau, {\it {The couplings of the Higgs boson and its CP
  properties from fits of the signal strengths and their ratios at the 7+8 TeV
  LHC}},  {\em Eur.Phys.J.} {\bf C73} (2013) 2512,
  [\href{http://xxx.lanl.gov/abs/1303.6591}{{\tt arXiv:1303.6591}}].

\bibitem{Cheung:2013kla}
K.~Cheung, J.~S. Lee, and P.-Y. Tseng, {\it {Higgs Precision (Higgcision) Era
  begins}},  {\em JHEP} {\bf 1305} (2013) 134,
  [\href{http://xxx.lanl.gov/abs/1302.3794}{{\tt arXiv:1302.3794}}].

\bibitem{Cheung:2014oaa}
K.~Cheung, J.~S. Lee, E.~Senaha, and P.-Y. Tseng, {\it {Confronting Higgcision
  with Electric Dipole Moments}},  {\em JHEP} {\bf 1406} (2014) 149,
  [\href{http://xxx.lanl.gov/abs/1403.4775}{{\tt arXiv:1403.4775}}].

\bibitem{Biswas:2012bd}
S.~Biswas, E.~Gabrielli, and B.~Mele, {\it {Single top and Higgs associated
  production as a probe of the Htt coupling sign at the LHC}},  {\em JHEP} {\bf
  1301} (2013) 088, [\href{http://xxx.lanl.gov/abs/1211.0499}{{\tt
  arXiv:1211.0499}}].

\bibitem{Biswas:2013xva}
S.~Biswas, E.~Gabrielli, F.~Margaroli, and B.~Mele, {\it {Direct constraints on
  the top-Higgs coupling from the 8 TeV LHC data}},  {\em JHEP} {\bf 1307}
  (2013) 073, [\href{http://xxx.lanl.gov/abs/1304.1822}{{\tt
  arXiv:1304.1822}}].

\bibitem{Ellis:2013yxa}
J.~Ellis, D.~S. Hwang, K.~Sakurai, and M.~Takeuchi, {\it {Disentangling
  Higgs-Top Couplings in Associated Production}},  {\em JHEP} {\bf 1404} (2014)
  004, [\href{http://xxx.lanl.gov/abs/1312.5736}{{\tt arXiv:1312.5736}}].

\bibitem{Yue:2014tya}
J.~Yue, {\it {$\mathcal{CP}$-violating top-Higgs coupling and top polarisation
  in the diphoton decay of the $thj$ channel at the LHC}},
  \href{http://xxx.lanl.gov/abs/1410.2701}{{\tt arXiv:1410.2701}}.

\bibitem{Chang:2014rfa}
J.~Chang, K.~Cheung, J.~S. Lee, and C.-T. Lu, {\it {Probing the Top-Yukawa
  Coupling in Associated Higgs production with a Single Top Quark}},  {\em
  JHEP} {\bf 1405} (2014) 062, [\href{http://xxx.lanl.gov/abs/1403.2053}{{\tt
  arXiv:1403.2053}}].

\bibitem{Kobakhidze:2014gqa}
A.~Kobakhidze, L.~Wu, and J.~Yue, {\it {Anomalous Top-Higgs Couplings and Top
  Polarisation in Single Top and Higgs Associated Production at the LHC}},
  {\em JHEP} {\bf 1410} (2014) 100,
  [\href{http://xxx.lanl.gov/abs/1406.1961}{{\tt arXiv:1406.1961}}].

\bibitem{Englert:2014pja}
C.~Englert and E.~Re, {\it {Bounding the top Yukawa coupling with
  Higgs-associated single-top production}},  {\em Phys.Rev.} {\bf D89} (2014),
  no.~7 073020, [\href{http://xxx.lanl.gov/abs/1402.0445}{{\tt
  arXiv:1402.0445}}].

\bibitem{CMS-PAS-HIG-14-001}
{\bf CMS Collaboration} Collaboration, {\it {Search for associated production
  of a single top quark and a Higgs boson in events where the Higgs boson
  decays to two photons at $\sqrt{s} = 8$~TeV}},  Tech. Rep.
  CMS-PAS-HIG-14-001, CERN, Geneva, 2014.

\bibitem{CMS-PAS-HIG-14-015}
{\bf CMS Collaboration} Collaboration, {\it {Search for H to bbbar in
  association with single top quarks as a test of Higgs couplings}},  Tech.
  Rep. CMS-PAS-HIG-14-015, CERN, Geneva, 2014.

\bibitem{CMS-PAS-HIG-14-026}
{\bf CMS Collaboration} Collaboration, {\it {Search for Associated Production
  of a Single Top Quark and a Higgs Boson in Leptonic Channels}},  Tech. Rep.
  CMS-PAS-HIG-14-026, CERN, Geneva, 2015.

\bibitem{Aad:2014lma}
{\bf ATLAS Collaboration} Collaboration, G.~Aad {\em et.~al.}, {\it {Search for
  $H \to \gamma\gamma$ produced in association with top quarks and constraints
  on the Yukawa coupling between the top quark and the Higgs boson using data
  taken at 7 TeV and 8 TeV with the ATLAS detector}},
  \href{http://xxx.lanl.gov/abs/1409.3122}{{\tt arXiv:1409.3122}}.

\bibitem{ATLAS-CONF-2014-011}
{\it {Search for the Standard Model Higgs boson produced in association with
  top quarks and decaying to $b\bar{b}$ in pp collisions at $\sqrt{s} =$ 8 TeV
  with the ATLAS detector at the LHC}},  Tech. Rep. ATLAS-CONF-2014-011, CERN,
  Geneva, Mar, 2014.

\bibitem{Khachatryan:2014qaa}
{\bf CMS Collaboration} Collaboration, V.~Khachatryan {\em et.~al.}, {\it
  {Search for the associated production of the Higgs boson with a top-quark
  pair}},  {\em JHEP} {\bf 1409} (2014) 087,
  [\href{http://xxx.lanl.gov/abs/1408.1682}{{\tt arXiv:1408.1682}}].

\bibitem{ATL-PHYS-PUB-2014-012}
{\it {HL-LHC projections for signal and background yield measurements of the
  $H\to\gamma\gamma$ when the Higgs boson is produced in association with $t$
  quarks, $W$ or $Z$ bosons}},  Tech. Rep. ATL-PHYS-PUB-2014-012, CERN, Geneva,
  Jul, 2014.

\bibitem{Plehn:2009rk}
T.~Plehn, G.~P. Salam, and M.~Spannowsky, {\it {Fat Jets for a Light Higgs}},
  {\em Phys.Rev.Lett.} {\bf 104} (2010) 111801,
  [\href{http://xxx.lanl.gov/abs/0910.5472}{{\tt arXiv:0910.5472}}].

\bibitem{Artoisenet:2013vfa}
P.~Artoisenet, P.~de~Aquino, F.~Maltoni, and O.~Mattelaer, {\it {Unravelling
  $t\overline{t}h$ via the Matrix Element Method}},  {\em Phys.Rev.Lett.} {\bf
  111} (2013), no.~9 091802, [\href{http://xxx.lanl.gov/abs/1304.6414}{{\tt
  arXiv:1304.6414}}].

\bibitem{Buckley:2013auc}
M.~R. Buckley, T.~Plehn, T.~Schell, and M.~Takeuchi, {\it {Buckets of Higgs and
  Tops}},  {\em JHEP} {\bf 1402} (2014) 130,
  [\href{http://xxx.lanl.gov/abs/1310.6034}{{\tt arXiv:1310.6034}}].

\bibitem{Maltoni:2002jr}
F.~Maltoni, D.~L. Rainwater, and S.~Willenbrock, {\it {Measuring the top quark
  Yukawa coupling at hadron colliders via $t\bar{t}H,H\to W^+W^-$}},  {\em
  Phys.Rev.} {\bf D66} (2002) 034022,
  [\href{http://xxx.lanl.gov/abs/hep-ph/0202205}{{\tt hep-ph/0202205}}].

\bibitem{Curtin:2013zua}
D.~Curtin, J.~Galloway, and J.~G. Wacker, {\it {Measuring the $t \bar th$
  coupling from same-sign dilepton $+2b$ measurements}},  {\em Phys.Rev.} {\bf
  D88} (2013), no.~9 093006, [\href{http://xxx.lanl.gov/abs/1306.5695}{{\tt
  arXiv:1306.5695}}].

\bibitem{Agrawal:2013owa}
P.~Agrawal, S.~Bandyopadhyay, and S.~P. Das, {\it {Multilepton Signatures of
  the Higgs Boson through its Production in Association with a Top-quark
  Pair}},  {\em Phys.Rev.} {\bf D88} (2013), no.~9 093008,
  [\href{http://xxx.lanl.gov/abs/1308.3043}{{\tt arXiv:1308.3043}}].

\bibitem{Stockinger:2006zn}
D.~Stockinger, {\it {The Muon Magnetic Moment and Supersymmetry}},  {\em
  J.Phys.} {\bf G34} (2007) R45--R92,
  [\href{http://xxx.lanl.gov/abs/hep-ph/0609168}{{\tt hep-ph/0609168}}].

\bibitem{Brod:2013cka}
J.~Brod, U.~Haisch, and J.~Zupan, {\it {Constraints on CP-violating Higgs
  couplings to the third generation}},  {\em JHEP} {\bf 1311} (2013) 180,
  [\href{http://xxx.lanl.gov/abs/1310.1385}{{\tt arXiv:1310.1385}}].

\bibitem{Arbey:2014msa}
A.~Arbey, J.~Ellis, R.~Godbole, and F.~Mahmoudi, {\it {Exploring CP Violation
  in the MSSM}},  \href{http://xxx.lanl.gov/abs/1410.4824}{{\tt
  arXiv:1410.4824}}.

\bibitem{CMS-PAS-HIG-14-009}
{\bf CMS Collaboration} Collaboration, {\it {Precise determination of the mass
  of the Higgs boson and studies of the compatibility of its couplings with the
  standard model}},  Tech. Rep. CMS-PAS-HIG-14-009, CERN, Geneva, 2014.

\bibitem{ATLAS-CONF-2014-009}
{\bf ATLAS} Collaboration, {\it Updated coupling measurements of the Higgs
  boson with the ATLAS detector using up to 25 fb$^{-1}$ of proton-proton
  collision data}, ATLAS-CONF-2014-009, ATLAS-COM-CONF-2014-013.

\bibitem{Aad:2015gra}
{\bf ATLAS Collaboration} Collaboration, G.~Aad {\em et.~al.}, {\it {Search for
  the Standard Model Higgs boson produced in association with top quarks and
  decaying into $b\bar{b}$ in pp collisions at $\sqrt{s}$ = 8 TeV with the
  ATLAS detector}},  \href{http://xxx.lanl.gov/abs/1503.0506}{{\tt
  arXiv:1503.0506}}.

\bibitem{ATLAS-CONF-2015-006}
{\it {Search for the associated production of the Higgs boson with a top quark
  pair in multi-lepton final states with the ATLAS detector}},  Tech. Rep.
  ATLAS-CONF-2015-006, CERN, Geneva, Mar, 2015.

\bibitem{Djouadi:2005gj}
A.~Djouadi, {\it {The Anatomy of electro-weak symmetry breaking. II. The Higgs
  bosons in the minimal supersymmetric model}},  {\em Phys.Rept.} {\bf 459}
  (2008) 1--241, [\href{http://xxx.lanl.gov/abs/hep-ph/0503173}{{\tt
  hep-ph/0503173}}].

\bibitem{Cacciapaglia:2012wb}
G.~Cacciapaglia, A.~Deandrea, G.~D. La~Rochelle, and J.-B. Flament, {\it {Higgs
  couplings beyond the Standard Model}},  {\em JHEP} {\bf 1303} (2013) 029,
  [\href{http://xxx.lanl.gov/abs/1210.8120}{{\tt arXiv:1210.8120}}].

\bibitem{Banerjee:2012xc}
S.~Banerjee, S.~Mukhopadhyay, and B.~Mukhopadhyaya, {\it {New Higgs
  interactions and recent data from the LHC and the Tevatron}},  {\em JHEP}
  {\bf 1210} (2012) 062, [\href{http://xxx.lanl.gov/abs/1207.3588}{{\tt
  arXiv:1207.3588}}].

\bibitem{Bhattacharyya:2012tj}
G.~Bhattacharyya, D.~Das, and P.~B. Pal, {\it {Modified Higgs couplings and
  unitarity violation}},  {\em Phys.Rev.} {\bf D87} (2013) 011702,
  [\href{http://xxx.lanl.gov/abs/1212.4651}{{\tt arXiv:1212.4651}}].

\bibitem{Choudhury:2012tk}
D.~Choudhury, R.~Islam, and A.~Kundu, {\it {Anomalous Higgs Couplings as a
  Window to New Physics}},  {\em Phys.Rev.} {\bf D88} (2013), no.~1 013014,
  [\href{http://xxx.lanl.gov/abs/1212.4652}{{\tt arXiv:1212.4652}}].

\bibitem{CMS-PAS-HIG-13-005}
{\bf CMS} Collaboration, {\it Combination of standard model Higgs boson
  searches and measurements of the properties of the new boson with a mass near
  125 GeV}, CMS-PAS-HIG-13-005.

\bibitem{ATLAS-CONF-2013-034}
{\bf ATLAS} Collaboration, {\it Combined coupling measurements of the
  Higgs-like boson with the ATLAS detector using up to 25 fb$^{-1}$ of
  proton-proton collision data}, ATLAS-CONF-2013-034, ATLAS-COM-CONF-2013-035.

\bibitem{Nishiwaki:2013cma}
K.~Nishiwaki, S.~Niyogi, and A.~Shivaji, {\it {$ttH$ Anomalous Coupling in
  Double Higgs Production}},  {\em JHEP} {\bf 1404} (2014) 011,
  [\href{http://xxx.lanl.gov/abs/1309.6907}{{\tt arXiv:1309.6907}}].

\bibitem{ATLAS_TDR}
{\bf ATLAS} Collaboration, {\it ATLAS detector and physics performance:
  Technical Design Report}, available on {\tt atlas.web.cern.ch}.

\bibitem{Ball:2007zza}
{\bf CMS Collaboration} Collaboration, G.~Bayatian {\em et.~al.}, {\it {CMS
  technical design report, volume II: Physics performance}},  {\em J.Phys.}
  {\bf G34} (2007) 995--1579.

\bibitem{Djouadi:2007ik}
{\bf ILC} Collaboration, G.~Aarons {\em et.~al.}, {\it {International Linear
  Collider Reference Design Report Volume 2: Physics at the ILC}},
  \href{http://xxx.lanl.gov/abs/0709.1893}{{\tt arXiv:0709.1893}}.

\bibitem{Djouadi:1992gp}
A.~Djouadi, J.~Kalinowski, and P.~Zerwas, {\it {Measuring the H t anti-t
  coupling in e+ e- collisions}},  {\em Mod.Phys.Lett.} {\bf A7} (1992)
  1765--1769.

\bibitem{Djouadi:1991tk}
A.~Djouadi, J.~Kalinowski, and P.~Zerwas, {\it {Higgs radiation off top quarks
  in high-energy e+ e- colliders}},  {\em Z.Phys.} {\bf C54} (1992) 255--262.

\bibitem{Grzadkowski:1999ye}
B.~Grzadkowski, J.~F. Gunion, and J.~Kalinowski, {\it {Finding the CP violating
  Higgs bosons at e+ e- colliders}},  {\em Phys.Rev.} {\bf D60} (1999) 075011,
  [\href{http://xxx.lanl.gov/abs/hep-ph/9902308}{{\tt hep-ph/9902308}}].

\bibitem{Dawson:1997im}
S.~Dawson and L.~Reina, {\it {QCD corrections to associated Higgs boson
  production}},  {\em Phys.Rev.} {\bf D57} (1998) 5851--5859,
  [\href{http://xxx.lanl.gov/abs/hep-ph/9712400}{{\tt hep-ph/9712400}}].

\bibitem{Dittmaier:1998dz}
S.~Dittmaier, M.~Kramer, Y.~Liao, M.~Spira, and P.~Zerwas, {\it {Higgs
  radiation off top quarks in e+ e- collisions}},  {\em Phys.Lett.} {\bf B441}
  (1998) 383--388, [\href{http://xxx.lanl.gov/abs/hep-ph/9808433}{{\tt
  hep-ph/9808433}}].

\bibitem{You:2003zq}
Y.~You, W.-G. Ma, H.~Chen, R.-Y. Zhang, S.~Yan-Bin, {\em et.~al.}, {\it
  {Electroweak radiative corrections to $e^+ e^- \rightarrow t \bar t h$ at
  linear colliders}},  {\em Phys.Lett.} {\bf B571} (2003) 85--91,
  [\href{http://xxx.lanl.gov/abs/hep-ph/0306036}{{\tt hep-ph/0306036}}].

\bibitem{Belanger:2003nm}
G.~Belanger, F.~Boudjema, J.~Fujimoto, T.~Ishikawa, T.~Kaneko, {\em et.~al.},
  {\it {Full O(alpha) electroweak and O(alpha(s)) corrections to $e^+ e^-
  \rightarrow t \bar t H$}},  {\em Phys.Lett.} {\bf B571} (2003) 163--172,
  [\href{http://xxx.lanl.gov/abs/hep-ph/0307029}{{\tt hep-ph/0307029}}].

\bibitem{BarShalom:1995jb}
S.~Bar-Shalom, D.~Atwood, G.~Eilam, R.~Mendel, and A.~Soni, {\it {Large tree
  level CP violation in $e^{+} e^{-} \to t \bar{t} H^0$ in the two Higgs
  doublet model}},  {\em Phys.Rev.} {\bf D53} (1996) 1162--1167,
  [\href{http://xxx.lanl.gov/abs/hep-ph/9508314}{{\tt hep-ph/9508314}}].

\bibitem{Atwood:2000tu}
D.~Atwood, S.~Bar-Shalom, G.~Eilam, and A.~Soni, {\it {CP violation in top
  physics}},  {\em Phys.Rept.} {\bf 347} (2001) 1--222,
  [\href{http://xxx.lanl.gov/abs/hep-ph/0006032}{{\tt hep-ph/0006032}}].

\bibitem{Gunion:1996vv}
J.~F. Gunion, B.~Grzadkowski, and X.-G. He, {\it {Determining the top -
  anti-top and Z Z couplings of a neutral Higgs boson of arbitrary CP nature at
  the NLC}},  {\em Phys.Rev.Lett.} {\bf 77} (1996) 5172--5175,
  [\href{http://xxx.lanl.gov/abs/hep-ph/9605326}{{\tt hep-ph/9605326}}].

\bibitem{Bhupal-Dev:2007is}
P.~Bhupal~Dev, A.~Djouadi, R.~Godbole, M.~Muhlleitner, and S.~Rindani, {\it
  {Determining the CP properties of the Higgs boson}},  {\em Phys.Rev.Lett.}
  {\bf 100} (2008) 051801, [\href{http://xxx.lanl.gov/abs/0707.2878}{{\tt
  arXiv:0707.2878}}].

\bibitem{Godbole:2011hw}
R.~Godbole, C.~Hangst, M.~Muhlleitner, S.~Rindani, and P.~Sharma, {\it
  {Model-independent analysis of Higgs spin and CP properties in the process
  $e^+ e^- \to t \bar t \Phi$}},  {\em Eur.Phys.J.} {\bf C71} (2011) 1681,
  [\href{http://xxx.lanl.gov/abs/1103.5404}{{\tt arXiv:1103.5404}}].

\bibitem{Godbole:2007uz}
R.~Godbole, P.~Bhupal~Dev, A.~Djouadi, M.~Muhlleitner, and S.~Rindani, {\it
  {Probing CP properties of the Higgs Boson via $e^+ e^- \to t \bar{t} \phi$}},
   {\em eConf} {\bf C0705302} (2007) TOP08,
  [\href{http://xxx.lanl.gov/abs/0710.2669}{{\tt arXiv:0710.2669}}].

\bibitem{Huang:2001ns}
C.-S. Huang and S.-h. Zhu, {\it {Heavy quark polarizations of $e^+ e^-
  \rightarrow q \bar q h$ in the general two Higgs doublet model}},  {\em
  Phys.Rev.} {\bf D65} (2002) 077702,
  [\href{http://xxx.lanl.gov/abs/hep-ph/0111280}{{\tt hep-ph/0111280}}].

\bibitem{Ananthanarayan:2013cia}
B.~Ananthanarayan, S.~K. Garg, J.~Lahiri, and P.~Poulose, {\it {Probing the
  indefinite CP nature of the Higgs boson through decay distributions in the
  process $e^+e^\to t\overline{t} \Phi$}},  {\em Phys.Rev.} {\bf D87} (2013),
  no.~11 114002, [\href{http://xxx.lanl.gov/abs/1304.4414}{{\tt
  arXiv:1304.4414}}].

\bibitem{Ananthanarayan:2014eea}
B.~Ananthanarayan, S.~K. Garg, C.~Kim, J.~Lahiri, and P.~Poulose, {\it {Top
  Yukawa coupling measurement with indefinite CP Higgs in $e^+e^-\to
  t\bar{t}\Phi$}},  {\em Phys.Rev.} {\bf D90} (2014) 014016,
  [\href{http://xxx.lanl.gov/abs/1405.6465}{{\tt arXiv:1405.6465}}].

\bibitem{Demartin:2014fia}
F.~Demartin, F.~Maltoni, K.~Mawatari, B.~Page, and M.~Zaro, {\it {Higgs
  characterisation at NLO in QCD: CP properties of the top-quark Yukawa
  interaction}},  {\em Eur.Phys.J.} {\bf C74} (2014), no.~9 3065,
  [\href{http://xxx.lanl.gov/abs/1407.5089}{{\tt arXiv:1407.5089}}].

\bibitem{Frederix:2011zi}
R.~Frederix, S.~Frixione, V.~Hirschi, F.~Maltoni, R.~Pittau, {\em et.~al.},
  {\it {Scalar and pseudoscalar Higgs production in association with a
  top-antitop pair}},  {\em Phys.Lett.} {\bf B701} (2011) 427--433,
  [\href{http://xxx.lanl.gov/abs/1104.5613}{{\tt arXiv:1104.5613}}].

\bibitem{Accomando:2006ga}
E.~Accomando, A.~Akeroyd, E.~Akhmetzyanova, J.~Albert, A.~Alves, {\em et.~al.},
  {\it {Workshop on CP Studies and Non-Standard Higgs Physics}},
  \href{http://xxx.lanl.gov/abs/hep-ph/0608079}{{\tt hep-ph/0608079}}.

\bibitem{Godbole:2004xe}
R.~Godbole, S.~Kraml, M.~Krawczyk, D.~Miller, P.~Niezurawski, {\em et.~al.},
  {\it {CP studies of the Higgs sector: A Contribution to the LHC / LC Study
  Group document}},  \href{http://xxx.lanl.gov/abs/hep-ph/0404024}{{\tt
  hep-ph/0404024}}.

\bibitem{Gunion:1996xu}
J.~F. Gunion and X.-G. He, {\it {Determining the CP nature of a neutral Higgs
  boson at the LHC}},  {\em Phys.Rev.Lett.} {\bf 76} (1996) 4468--4471,
  [\href{http://xxx.lanl.gov/abs/hep-ph/9602226}{{\tt hep-ph/9602226}}].

\bibitem{He:2014xla}
X.-G. He, G.-N. Li, and Y.-J. Zheng, {\it {Probing Higgs Boson CP Properties
  with $t\bar{t}H$ at the LHC and the 100 TeV pp Collider}},
  \href{http://xxx.lanl.gov/abs/1501.0001}{{\tt arXiv:1501.0001}}.

\bibitem{Khatibi:2014bsa}
S.~Khatibi and M.~M. Najafabadi, {\it {Exploring the Anomalous Higgs-top
  Couplings}},  {\em Phys.Rev.} {\bf D90} (2014), no.~7 074014,
  [\href{http://xxx.lanl.gov/abs/1409.6553}{{\tt arXiv:1409.6553}}].

\bibitem{Atwood:1991ka}
D.~Atwood and A.~Soni, {\it {Analysis for magnetic moment and electric dipole
  moment form-factors of the top quark via e+ e- $\to$ t anti-t}},  {\em
  Phys.Rev.} {\bf D45} (1992) 2405--2413.

\bibitem{Alwall:2014hca}
J.~Alwall, R.~Frederix, S.~Frixione, V.~Hirschi, F.~Maltoni, {\em et.~al.},
  {\it {The automated computation of tree-level and next-to-leading order
  differential cross sections, and their matching to parton shower
  simulations}},  {\em JHEP} {\bf 1407} (2014) 079,
  [\href{http://xxx.lanl.gov/abs/1405.0301}{{\tt arXiv:1405.0301}}].

\bibitem{Baglio:2012et}
J.~Baglio, A.~Djouadi, and R.~Godbole, {\it {The apparent excess in the Higgs
  to di-photon rate at the LHC: New Physics or QCD uncertainties?}},  {\em
  Phys.Lett.} {\bf B716} (2012) 203--207,
  [\href{http://xxx.lanl.gov/abs/1207.1451}{{\tt arXiv:1207.1451}}].

\bibitem{Boudjema:2009fz}
F.~Boudjema and R.~K. Singh, {\it {A Model independent spin analysis of
  fundamental particles using azimuthal asymmetries}},  {\em JHEP} {\bf 0907}
  (2009) 028, [\href{http://xxx.lanl.gov/abs/0903.4705}{{\tt
  arXiv:0903.4705}}].

\bibitem{Buckley:2007th}
M.~R. Buckley, H.~Murayama, W.~Klemm, and V.~Rentala, {\it {Discriminating spin
  through quantum interference}},  {\em Phys.Rev.} {\bf D78} (2008) 014028,
  [\href{http://xxx.lanl.gov/abs/0711.0364}{{\tt arXiv:0711.0364}}].

\bibitem{Godbole:2006tq}
R.~M. Godbole, S.~D. Rindani, and R.~K. Singh, {\it {Lepton distribution as a
  probe of new physics in production and decay of the t quark and its
  polarization}},  {\em JHEP} {\bf 0612} (2006) 021,
  [\href{http://xxx.lanl.gov/abs/hep-ph/0605100}{{\tt hep-ph/0605100}}].

\bibitem{Murayama:1992gi}
H.~Murayama, I.~Watanabe, and K.~Hagiwara, {\it {HELAS: HELicity amplitude
  subroutines for Feynman diagram evaluations}}, .

\bibitem{Mahlon:2010gw}
G.~Mahlon and S.~J. Parke, {\it {Spin Correlation Effects in Top Quark Pair
  Production at the LHC}},  {\em Phys.Rev.} {\bf D81} (2010) 074024,
  [\href{http://xxx.lanl.gov/abs/1001.3422}{{\tt arXiv:1001.3422}}].

\bibitem{Mahlon:1995zn}
G.~Mahlon and S.~J. Parke, {\it {Angular correlations in top quark pair
  production and decay at hadron colliders}},  {\em Phys.Rev.} {\bf D53} (1996)
  4886--4896, [\href{http://xxx.lanl.gov/abs/hep-ph/9512264}{{\tt
  hep-ph/9512264}}].

\bibitem{ATLAS:2012ao}
{\bf ATLAS Collaboration} Collaboration, G.~Aad {\em et.~al.}, {\it
  {Observation of spin correlation in $t \bar{t}$ events from pp collisions at
  sqrt(s) = 7 TeV using the ATLAS detector}},  {\em Phys.Rev.Lett.} {\bf 108}
  (2012) 212001, [\href{http://xxx.lanl.gov/abs/1203.4081}{{\tt
  arXiv:1203.4081}}].

\bibitem{CMS-PAS-TOP-12-004}
{\bf CMS} Collaboration, {\it Measurement of Spin Correlations in ttbar
  production}, CMS-PAS-TOP-12-004.

\bibitem{Godbole:2002qu}
R.~M. Godbole, S.~D. Rindani, and R.~K. Singh, {\it {Study of CP property of
  the Higgs at a photon collider using $\gamma \gamma \rightarrow t \bar t
  \rightarrow lX$}},  {\em Phys.Rev.} {\bf D67} (2003) 095009,
  [\href{http://xxx.lanl.gov/abs/hep-ph/0211136}{{\tt hep-ph/0211136}}].

\bibitem{Heinemeyer:2013tqa}
{\bf LHC Higgs Cross Section Working Group} Collaboration, S.~Heinemeyer {\em
  et.~al.}, {\it {Handbook of LHC Higgs Cross Sections: 3. Higgs Properties}},
  \href{http://xxx.lanl.gov/abs/1307.1347}{{\tt arXiv:1307.1347}}.

\bibitem{Biswas:2014hwa}
S.~Biswas, R.~Frederix, E.~Gabrielli, and B.~Mele, {\it {Enhancing the
  $t\bar{t}H$ signal through top-quark spin polarization effects at the LHC}},
  {\em JHEP} {\bf 1407} (2014) 020,
  [\href{http://xxx.lanl.gov/abs/1403.1790}{{\tt arXiv:1403.1790}}].

\bibitem{Brooijmans:2014eja}
G.~Brooijmans, R.~Contino, B.~Fuks, F.~Moortgat, P.~Richardson, {\em et.~al.},
  {\it {Les Houches 2013: Physics at TeV Colliders: New Physics Working Group
  Report}},  \href{http://xxx.lanl.gov/abs/1405.1617}{{\tt arXiv:1405.1617}}.

\bibitem{Aad:2013uza}
{\bf ATLAS} Collaboration, G.~Aad {\em et.~al.}, {\it {Measurement of the top
  quark charge in $pp$ collisions at $\sqrt{s} =$ 7 TeV with the ATLAS
  detector}},  {\em JHEP} {\bf 1311} (2013) 031,
  [\href{http://xxx.lanl.gov/abs/1307.4568}{{\tt arXiv:1307.4568}}].

\bibitem{Nachman:2014qma}
{\bf ATLAS Collaboration} Collaboration, B.~Nachman, {\it {Jet Charge with the
  ATLAS Detector using $\sqrt{s}=8$ TeV $pp$ Collision Data}},
  \href{http://xxx.lanl.gov/abs/1409.0318}{{\tt arXiv:1409.0318}}.

\bibitem{CMS:2012wcp}
{\bf CMS Collaboration} Collaboration, C.~Collaboration, {\it {Constraints on
  the Top-Quark Charge from Top-Pair Events}}, .

\bibitem{Krohn:2012fg}
D.~Krohn, M.~D. Schwartz, T.~Lin, and W.~J. Waalewijn, {\it {Jet Charge at the
  LHC}},  {\em Phys.Rev.Lett.} {\bf 110} (2013), no.~21 212001,
  [\href{http://xxx.lanl.gov/abs/1209.2421}{{\tt arXiv:1209.2421}}].

\bibitem{Aad:2014eha}
{\bf ATLAS Collaboration} Collaboration, G.~Aad {\em et.~al.}, {\it
  {Measurement of Higgs boson production in the diphoton decay channel in $pp$
  collisions at center-of-mass energies of 7 and 8 TeV with the ATLAS
  detector}},  \href{http://xxx.lanl.gov/abs/1408.7084}{{\tt arXiv:1408.7084}}.

\bibitem{Khachatryan:2014ira}
{\bf CMS Collaboration} Collaboration, V.~Khachatryan {\em et.~al.}, {\it
  {Observation of the diphoton decay of the Higgs boson and measurement of its
  properties}},  \href{http://xxx.lanl.gov/abs/1407.0558}{{\tt
  arXiv:1407.0558}}.

\bibitem{Group:2012zca}
{\bf Tevatron New Physics Higgs Working Group, CDF Collaboration, D0
  Collaboration} Collaboration, {\it {Updated Combination of CDF and D0
  Searches for Standard Model Higgs Boson Production with up to 10.0 fb$^{-1}$
  of Data}},  \href{http://xxx.lanl.gov/abs/1207.0449}{{\tt arXiv:1207.0449}}.

\bibitem{Aad:2014aba}
{\bf ATLAS Collaboration} Collaboration, G.~Aad {\em et.~al.}, {\it
  {Measurement of the Higgs boson mass from the $H\rightarrow \gamma\gamma$ and
  $H \rightarrow ZZ^{*} \rightarrow 4\ell$ channels with the ATLAS detector
  using 25 fb$^{-1}$ of $pp$ collision data}},  {\em Phys.Rev.} {\bf D90}
  (2014) 052004, [\href{http://xxx.lanl.gov/abs/1406.3827}{{\tt
  arXiv:1406.3827}}].

\bibitem{Aad:2014eva}
{\bf ATLAS Collaboration} Collaboration, G.~Aad {\em et.~al.}, {\it
  {Measurements of Higgs boson production and couplings in the four-lepton
  channel in pp collisions at center-of-mass energies of 7 and 8 TeV with the
  ATLAS detector}},  \href{http://xxx.lanl.gov/abs/1408.5191}{{\tt
  arXiv:1408.5191}}.

\bibitem{ATLAS-CONF-2014-060}
{\it {Observation and measurement of Higgs boson decays to $WW^\ast$ with ATLAS
  at the LHC}},  Tech. Rep. ATLAS-CONF-2014-060, CERN, Geneva, Oct, 2014.

\bibitem{Aad:2014xzb}
{\bf ATLAS Collaboration} Collaboration, G.~Aad {\em et.~al.}, {\it {Search for
  the $b\bar{b}$ decay of the Standard Model Higgs boson in associated $(W/Z)H$
  production with the ATLAS detector}},
  \href{http://xxx.lanl.gov/abs/1409.6212}{{\tt arXiv:1409.6212}}.

\bibitem{Chatrchyan:2013zna}
{\bf CMS Collaboration} Collaboration, S.~Chatrchyan {\em et.~al.}, {\it
  {Search for the standard model Higgs boson produced in association with a W
  or a Z boson and decaying to bottom quarks}},  {\em Phys.Rev.} {\bf D89}
  (2014), no.~1 012003, [\href{http://xxx.lanl.gov/abs/1310.3687}{{\tt
  arXiv:1310.3687}}].

\bibitem{ATLAS-CONF-2014-061}
{\it {Evidence for Higgs boson Yukawa couplings in the $H\to\tau\tau$ decay
  mode with the ATLAS detector}},  Tech. Rep. ATLAS-CONF-2014-061, CERN,
  Geneva, Oct, 2014.

\bibitem{Chatrchyan:2014nva}
{\bf CMS Collaboration} Collaboration, S.~Chatrchyan {\em et.~al.}, {\it
  {Evidence for the 125 GeV Higgs boson decaying to a pair of $\tau$ leptons}},
   {\em JHEP} {\bf 1405} (2014) 104,
  [\href{http://xxx.lanl.gov/abs/1401.5041}{{\tt arXiv:1401.5041}}].

\bibitem{Spira:1995rr}
M.~Spira, A.~Djouadi, D.~Graudenz, and P.~Zerwas, {\it {Higgs boson production
  at the LHC}},  {\em Nucl.Phys.} {\bf B453} (1995) 17--82,
  [\href{http://xxx.lanl.gov/abs/hep-ph/9504378}{{\tt hep-ph/9504378}}].

\bibitem{Djouadi:1993ji}
A.~Djouadi, M.~Spira, and P.~Zerwas, {\it {Two photon decay widths of Higgs
  particles}},  {\em Phys.Lett.} {\bf B311} (1993) 255--260,
  [\href{http://xxx.lanl.gov/abs/hep-ph/9305335}{{\tt hep-ph/9305335}}].

\end{thebibliography}\endgroup

\end{document}